\documentclass{article}

\usepackage{arxiv}

\usepackage[utf8]{inputenc} % allow utf-8 input
\usepackage[T1]{fontenc}    % use 8-bit T1 fonts
\usepackage{hyperref}       % hyperlinks
\usepackage{url}            % simple URL typesetting
\usepackage{booktabs}       % professional-quality tables
\usepackage{amsfonts}       % blackboard math symbols
\usepackage{nicefrac}       % compact symbols for 1/2, etc.
\usepackage{microtype}      % microtypography
\usepackage{lipsum}

\usepackage{colortbl}
\usepackage{amsmath,amssymb}
\usepackage{bm}
\usepackage{lastpage,fancyhdr,graphicx}
\usepackage{listings}
\usepackage{xcolor}
\usepackage{multirow}
\title{Physics-informed Neural Networks for Solving Inverse 
Problems of Nonlinear Biot's Equations: Batch Training}

\author{
  Teeratorn Kadeethum \\
  Department of Applied Mathematics and Computer Science\\
  Technical University of Denmark\\
   Lyngby, Denmark \\
  \texttt{teekad@dtu.dk} \\
   \And
 Thomas M Jørgensen \\
  Department of Applied Mathematics and Computer Science\\
  Technical University of Denmark\\
   Lyngby, Denmark \\
  \texttt{tmjq@dtu.dk} \\
   \And
 Hamidreza M Nick \\
  The Danish Hydrocarbon Research and Technology Centre\\
  Technical University of Denmark\\
   Lyngby, Denmark \\
  \texttt{hamid@dtu.dk} \\
}

\begin{document}
\maketitle

\begin{abstract}
In biomedical engineering, earthquake prediction, and underground energy harvesting, it is crucial to indirectly estimate the physical properties of porous media since the direct measurement of those are usually impractical/prohibitive. Here we apply the physics-informed neural networks to solve the inverse problem with regard to the nonlinear Biot's equations. Specifically, we consider batch training and explore the effect of different batch sizes. The results show that training with small batch sizes, i.e., a few examples per batch, provides better approximations (lower percentage error) of the physical parameters than using large batches or the full batch. The increased accuracy of the physical parameters, comes at the cost of longer training time. Specifically, we find the size should not be too small since a very small batch size requires a very long training time without a corresponding improvement in estimation accuracy. We find that a batch size of 8 or 32 is a good compromise, which is also robust to additive noise in the data. The learning rate also plays an important role and should be used as a hyperparameter.
\end{abstract}

% keywords can be removed
% keywords can be removed

\section{Introduction}
The volumetric displacement of a porous medium caused by the changes in fluid pressure inside the pore spaces is essential for many applications including groundwater flow, underground heat mining, fossil fuel production, earthquake mechanics, and biomedical engineering \cite{bisdom2016geometrically,juanes2016were, kadeethum2020well, kadeethum2019investigation,nick2013reactive,vinje2018fluid,lee2016phase}. Such volumetric deformation may impact the hydraulic storability and permeability of porous material, which in turn, influences the fluid flow field. This coupling between fluid flow and solid deformation can be captured through the application of Biot's equations of poroelasticity \cite{biot1941general,biot1957elastic}. The Biot's equations can be solved analytically for simple cases \cite{terzaghi1951theoretical,wang2017theory}, using finite volume discretization \cite{nordbotten2014cell,sokolova2019multiscale}, or more commonly by applying finite element methods \cite{choo2018enriched, Haga2012, Kadeethum2019ARMA,murad2013new,SALIMZADEH2018212,wheeler2014coupling} for complex systems. The last two methods, however, require significant computational resources and, therefore, may not be suitable to handle an inverse problem \cite{hansen2010discrete,hesthaven2016certified}.

In the past decades, deep learning has been successfully applied to many applications \cite{alipanahi2015predicting,krizhevsky2012imagenet,shen2017deep} because of its capability to handle highly nonlinear problems \cite{lecun2015deep}. This technique, in general, requires a significantly large data set to reach a reasonable accuracy \cite{ahmed2015improved}, hindering its applicability to many scientific and industrial problems \cite{raissi2017inferring}. Recently, the idea of encoding the physical laws into the architectures and loss functions of the deep neural network – so-called Physics Informed Neural Network (PINN) –  has been successfully applied to computational fluid mechanics problems \cite{raissi2019physics,wang2017physics,xiao2016quantifying} and the coupled solid and fluid problem \cite{kadeethum2020pinn}. The published results illustrate that by incorporating the physical laws, the neural network can provide good approximations with a limited data set.

Developing a fast and reliable method for parameter estimation in the case of Biot’s equations is desirable because the physical properties of the porous media are nontrivial and costly to measure as the media of interest usually locate in areas difficult to access such as inside living organisms or deep underground \cite{matthai2009upscaling,ruiz2019modelling}. Therefore, a non-intrusive or indirect measurement combined with an inverse model is highly beneficial \cite{horne1995modern,rogers1989}.

Since the coupled fluid and solid mechanics process is highly nonlinear \cite{choo2018enriched,Kadeethum2019ARMA,ruiz2019modelling}, it is a suitable problem to consider in the context of deep learning algorithms. Our previous study has shown that the PINN model could solve the nonlinear Biot’s equations for both forward and inverse modeling using full-batch training and limited-memory Broyden–Fletcher–Goldfarb–Shanno (L-BFGS)  algorithm as a minimization algorithm \cite{kadeethum2020pinn}. This study extends our previous model to the batch training focusing on the inverse problem of the nonlinear Biot’s equations because the batch training may provide more accurate results since it might avoid local minimum \cite{keskar2016largebatch}.

\section{Governing equations}

We consider an initial-boundary value problem of a poromechanical process, which couples solid deformation and fluid flow problems. Let $\Omega \subset \mathbb{R}^{d}(d \in\{2,3\})$ be the domain of interest in $d$-dimensional space that is bounded by boundaries, $\partial \Omega$. Note that we only focus on $d=2$ in this paper. The $\partial \Omega$ can be decomposed to displacement and traction boundaries, $\partial \Omega_{u}$ and $\partial \Omega_{t}$, respectively, for the solid deformation problem. For the fluid flow problem, $\partial \Omega$ is decomposed to pressure and flux boundaries, $\partial \Omega_{p}$ and $\partial \Omega_{q}$, respectively. The time domain is denoted by $\mathbb{T}=(0, \mathrm{T}]$ with $T>0$. 

The coupling between the fluid flow and solid deformation can be captured through the application of Biot's equations of poroelasticity, which is composed of two balance equations \cite{biot1941general,biot1957elastic}. The first equation is the linear momentum balance equation, as shown below:

\begin{equation}\label{eq:linear_balance}
\nabla \cdot \left( 2 \mu_{l} \bm{\varepsilon}(\bm{u})+\lambda_{l}  \boldsymbol{u} \bm{I} \right) +\alpha \nabla \cdot p \bm{I} -\rho \mathbf{g} = \bm{f} \text { in } \Omega \times \mathbb{T},
\end{equation}

\begin{equation}\label{eq:linear_balance_D}
\boldsymbol{u}=\boldsymbol{u}_{D} \text { on } \partial \Omega_{u} \times \mathbb{T},
\end{equation}

\begin{equation}\label{eq:linear_balance_Omega_t}
\boldsymbol{\sigma} \cdot \bm{n}=\bm{\sigma_{D}} \text { on } \partial \Omega_{\sigma} \times \mathbb{T},
\end{equation}

\begin{equation}
\bm{u}=\bm{u}_{0} \text { in } \Omega \text { at } t = 0,
\end{equation}

\noindent
where $\lambda_{l}$ and $\mu_{l}$ are Lam\'e constants, $\bm{u}$ is displacement, $\bm{I}$ is the second-order identity tensor, $p$ is fluid pressure, $\rho$ is the fluid density, $\mathbf{g}$ is a gravitational vector, and $\bm{f}$ is a sink/source term for this momentum balance equation. The $\rho \mathbf{g}$ term, or in other words, the body force, is neglected in this study. $\boldsymbol{u}_{D}$ is the prescribed displacement at the boundaries. $\alpha$ denotes Biot’s coefficient defined as \cite{jaeger2009fundamentals}:

\begin{equation}\alpha:=1-\frac{K}{K_{s}},\end{equation}

\noindent
where $K$ is the bulk modulus of a rock matrix, $K_{s}$ is bulk rock matrix material (e.g., solid grains). $\bm{\sigma}$ is the total stress defined as:

\begin{equation}\bm{\sigma}:=2 \mu_{l} \bm{\varepsilon}+\lambda_{l} \boldsymbol{u} \boldsymbol{I},\end{equation}

and $\bm{\sigma_{D}}$ is the prescribed traction at the boundaries. Assuming infinitesimal displacements, the strain, $\bm{\varepsilon}(\bm{u})$, is defined as:

\begin{equation}
\bm{\varepsilon}(\bm{u}) :=\frac{1}{2}\left(\nabla \boldsymbol{u}+\nabla^{T} \boldsymbol{u}\right).
\end{equation}

The second equation is the mass balance equation, which is given as:

\begin{equation} \label{eq:mass_balance}
\left(\phi c_{f}+\frac{\alpha-\phi}{K_{s}}\right) \frac{\partial p}{\partial t}+ \alpha \frac{\partial \nabla \cdot \boldsymbol{u}}{\partial t}-\nabla \cdot \mathcal{N}[\bm{\kappa}](\nabla p-\rho \mathbf{g})=g \text { in } \Omega \times \mathbb{T},
\end{equation}

\begin{equation} \label{eq:mass_balance_D}
p=p_{D} \text { on } \partial \Omega_{p} \times \mathbb{T},
\end{equation}

\begin{equation} \label{eq:mass_balance_N}
-\mathcal{N}[\bm{\kappa}](\nabla p-\rho \mathbf{g}) \cdot \boldsymbol{n}=q_{D} \text { on } \partial \Omega_{q} \times \mathbb{T},
\end{equation}

\begin{equation} \label{eq:mass_balance_I}
p=p_{0} \text { in } \Omega \text { at } t = 0,
\end{equation}

\noindent
where $\phi$ is initial porosity and remains constant throughout the simulation (the volumetric deformation is represented by $\partial \nabla \cdot \boldsymbol{u} / \partial t$), $c_f$ is fluid compressibility, $g$ is sink/source, $p_D$ and $q_D$ are specified pressure and flux, respectively, $\bm{\kappa}$ is hydraulic conductivity, and $\mathcal{N}[\cdot]$ is a nonlinear operator. In this paper, we simplify our problem by taking $\Omega=[0,1]^{2}$, $\mathbb{T}=[0,1]$, and choose the exact solution in $\Omega$ as:

\begin{equation}\boldsymbol{u}(x, y, t):=\left[\begin{array}{l}
u \\
v
\end{array}\right]=\left[\begin{array}{l}
\sin (x+y+t) \\
\cos (x+y+t)
\end{array}\right],\label{eq:biot_displacement_exact}\end{equation}

for the displacement variable where $u$ and $v$ are displacements in $x$- and $y$-direction, respectively. Note that because we focus on the 2-Dimensional domain, $u(x, y, t)$ is composed of two spatial components. For the pressure variable, we choose

\begin{equation}p(x, y, t):=e^{(x+y+t)}.\label{eq:biot_pressure_exact}\end{equation}

Here $x$, $y$, and $t$ represent points in $x$-, $y$-direction, and time domain, respectively. The choice of the $\mathcal{N}[\bm{\kappa}]$ function is selected to represent the change in a volumetric strain that affects the porous media conductivity, and it is adapted from \cite{abou2013petroleum,Du2007}. $\mathcal{N}[\bm{\kappa}]$ then takes the form:

\begin{equation}\mathcal{N}[\bm{\kappa}]:=\bm{\kappa}_{0} e^{\varepsilon_{v}}\end{equation}

\noindent
where $\bm{\kappa}_{0}$ represent initial rock matrix conductivity. We assume  $\bm{\kappa}_{0}$ to be a scalar in this paper, i.e.,  $\bm{\kappa}_{0}$ =  ${\kappa}_{0}$ . $\varepsilon_{v}$ is
the total volumetric strain defined as:

\begin{equation}\varepsilon_{v}:=\operatorname{tr}(\varepsilon).\end{equation}

\noindent
All the physical constants are set to 1.0; and subsequently, $\bm{f}$ is chosen as:

\begin{equation}\boldsymbol{f}(x, y, t):=\left[\begin{array}{l}
f_{u}(x, y, t) \\
f_{v}(x, y, t)
\end{array}\right],\end{equation}

\noindent
where

\begin{equation}\label{eq:biot_displacement_source1}
\begin{aligned}
f_u(x,y,t) & := -4.0\sin(x+y+t)-2.0\cos(x+y+t) -  e^{(x+y+t)},
\end{aligned}
\end{equation}

\noindent
and

\begin{equation}\label{eq:biot_displacement_source2}
\begin{aligned}
f_v(x,y,t) & := -4.0\cos(x+y+t)-2.0\sin(x+y+t) -  e^{(x+y+t)},
\end{aligned}
\end{equation}

\noindent
for the momentum balance equation, Eq~(\ref{eq:linear_balance}). The source term of the mass balance equation, Eq~(\ref{eq:mass_balance}), $g$ is chosen as:

\begin{equation}  \label{eq:biot_mass_source}
\begin{aligned}
g(x,y,t) &:= \left( \cos \left( x+y+t \right) +\sin \left( x+y+t \right) -1
 \right) {{\rm e}^{\cos \left( x+y+t \right) -\sin \left( x+y+t
 \right) +x+y+t}} \\
 &-\cos \left( x+y+t \right) +{{\rm e}^{x+y+t}}-\sin
 \left( x+y+t \right), 
\end{aligned}
\end{equation}

\noindent
to satisfy the exact solution. Furthermore, the boundary conditions and initial conditions are applied using Eqs~(\ref{eq:biot_displacement_exact}) and (\ref{eq:biot_pressure_exact}). 

We generate the exact solution points, Eqs~(\ref{eq:biot_displacement_exact}) and (\ref{eq:biot_pressure_exact}), based on a rectangular mesh $\Omega=[0,1]^{2}$ with 99 equidistant intervals in both $x$- and $y$-direction, i.e., $\Delta x$ = $\Delta y$. Using 49 equidistant temporal intervals, in total, we have 500000 examples. We draw n training examples randomly. Subsequently, we split the remaining examples equally for validation and test sets. Assuming we have 500000 solution points for the sake of illustration, we use 100 examples to train the model; we then have 249950 examples for both the validation and the test sets. We now formulate the two governing equations, Eqs. (\ref{eq:linear_balance}) and (\ref{eq:mass_balance}), in a parametrized form \cite{hesthaven2016certified} which will serve as the physics-informed constraints \cite{kadeethum2020pinn} to our neural network:

\begin{equation}\label{eq:biot_mom_inv}
\Pi_{\bm{u}} = \nabla \cdot \left [ 2 \theta_1 \bm{\varepsilon}(\bm{u})+\theta_2  \boldsymbol{u} \bm{I} \right ] + \theta_3 \nabla \cdot p \bm{I} - \bm{f} \text { in } \Omega \times \mathbb{T},
\end{equation}

\begin{equation} \label{eq:biot_mass_inv}
\Pi_p = \theta_4 \frac{\partial p}{\partial t}+\theta_3 \frac{\partial \nabla \cdot \boldsymbol{u}}{\partial t}- \theta_5 \nabla \cdot e^{\varepsilon_{v}}(\nabla p-\rho \mathbf{g}) - g \text { in } \Omega \times \mathbb{T},
\end{equation}

\noindent
where

\begin{equation} \label{eq:biot_theta}
\begin{aligned}
\theta_1 = \mu_l \text{,} \quad \theta_2 = \lambda_l \text{,} \quad
\theta_3 = \alpha \text{,} \quad
\theta_4 = \phi c_{f}+\frac{\alpha-\phi}{K_{s}}  \text{, } \text{and} \quad 
\theta_5 = {\kappa}_0.
\end{aligned}
\end{equation}

\section{Physics-informed neural networks}

\begin{figure}[!ht]
   \centering
        \includegraphics[width=9.0cm, height=9.0cm]{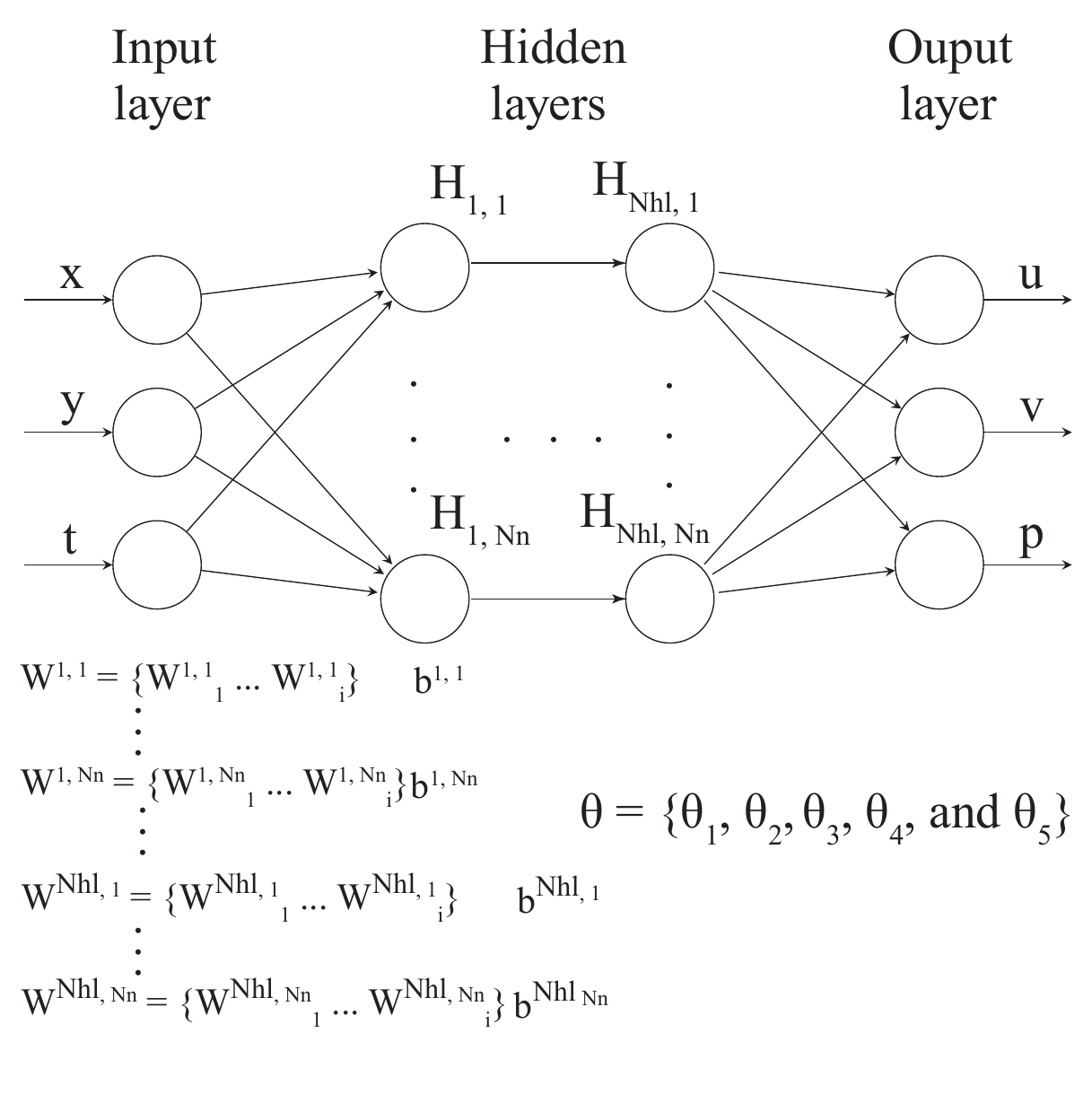}
   \caption{Neural network architecture used for solving an inverse problem of the nonlinear Biot's equations. There are three inputs, $x$, $y$, and $t$, and three outputs, $u$, $v$, and $p$. $N_{h l}$ refers to the number of hidden layers and each hidden layer is composed of $N_n$ neurons. Each neuron (e.g., $H_{1,1} \ldots H_{1, N_{n}}$) is connected to the nodes of the previous layer with adjustable weights (W) and also has an adjustable bias (b). In the inverse problem, the $\theta$ parameters act as adjustable variables along with  W and b.}
   \label{fig:des_for_inv}
\end{figure}

This paper is an extension of our previous work \cite{kadeethum2020pinn}. Therefore, due to the limited
space, we give only a short description of the physics-informed neural network architecture and its loss function. Detailed information can be found in \cite{kadeethum2020pinn}. Our network architecture used in this problem is illustrated in Fig \ref{fig:des_for_inv}. The main idea of solving inverse modeling using PINN is that we want to estimate a set of $\theta$, Eq. \ref{eq:biot_theta}, from a known set of examples/measurements relating the input space $x$, $y$, $t$, to the output space $u$, $v$, and $p$. Specifically, we train a neural network to predict $u$, $v$, and $p$ from given values of $x$, $y$, and $t$, and during training, the set of $\theta$ parameters are learned along with the weights (W) and biases (b) of the network itself. In other words, the reasoning behind solving the inverse problem is that we expect the unknown $\theta$ to converge towards their true values during training.

\begin{figure}[!ht]
   \centering
        \includegraphics[width=9.0cm, height=8.0cm]{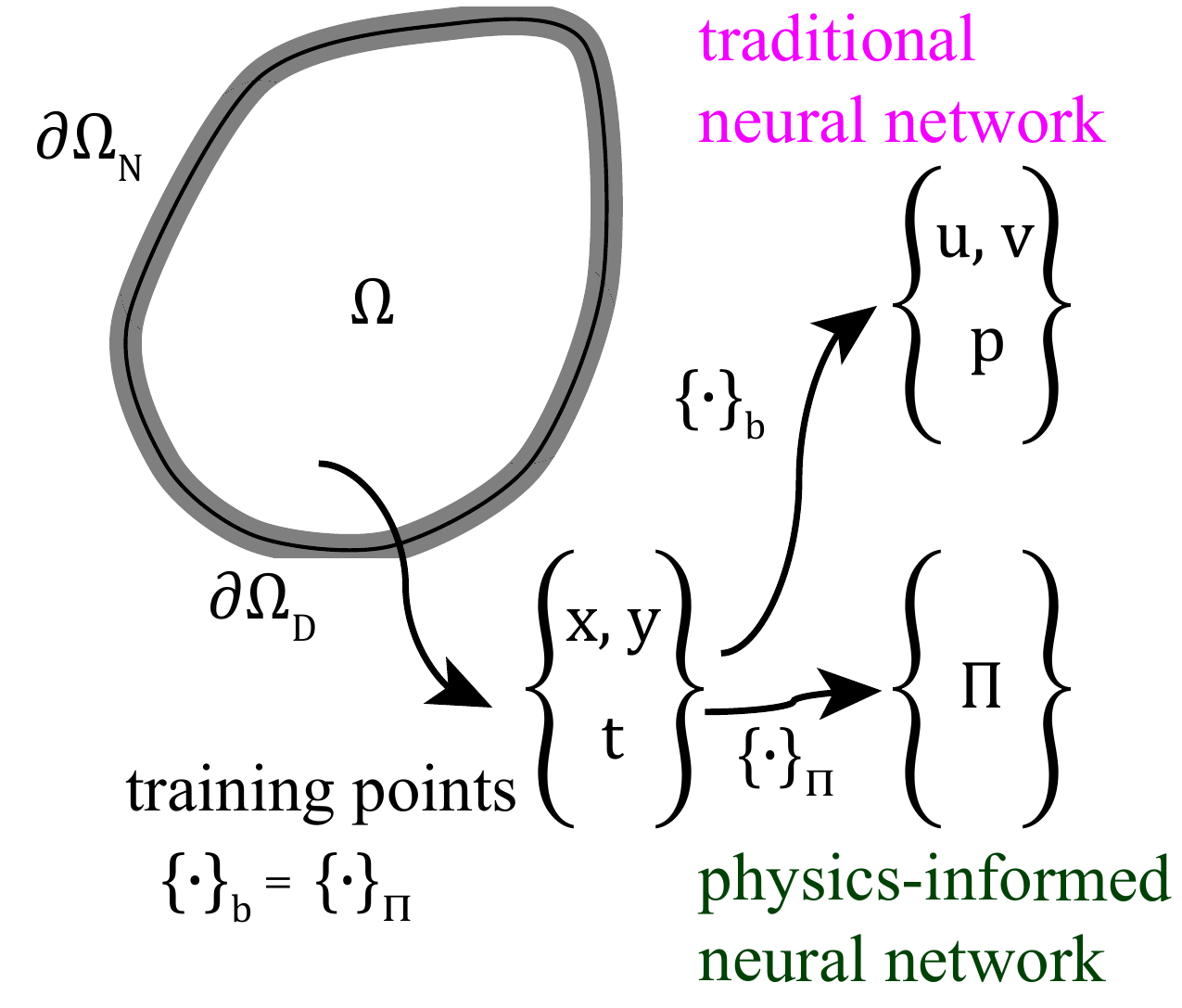}
   \caption{Illustration of the input space for training the PINN
when applied to the inverse problem.}
   \label{fig:explain_inv_nn}
\end{figure}

The loss function applied with the PINN scheme is composed of two parts (here we use a mean squared error - $MSE$ as the metric). The error on the training data ($MSE_b$) and the mean square value of the regularization term given by the physics-informed function ($M S E_{\Pi}$):

\begin{equation}M S E=M S E_{b}+M S E_{\Pi}, \label{eq:loss_pinn_batch}\end{equation}

\noindent
where

\begin{equation}M S E_{\Pi}=M S E_{\Pi_{u}}+M S E_{\Pi_{p}},\end{equation}

\noindent
where

\begin{equation}
M S E_{b}=\frac{1}{N_{b}} \sum_{i=1}^{N_{b}}\left[\left|u\left(x_{b}^{i}, y_{b}^{i}, t_{b}^{i}\right)-u^{i}\right|^{2} +\left|v\left(x_{b}^{i}, y_{b}^{i}, t_{b}^{i}\right)-v^{i}\right|^{2}
+\left|p\left(x_{b}^{i}, y_{b}^{i}, t_{b}^{i}\right)-p^{i}\right|^{2}\right],
\end{equation}

\begin{equation}\operatorname{MSE}_{\Pi_{u}}=\frac{1}{N_{\Pi}} \sum_{i=1}^{N_{\Pi}}\left|\Pi_{u}\left(x_{\Pi}^{i}, y_{\Pi}^{i}, t_{\Pi}^{i}\right)\right|^{2},\end{equation}

\noindent
and

\begin{equation}M S E_{\Pi_{p}}=\frac{1}{N_{\Pi}} \sum_{i=1}^{N_{\Pi}}\left|\Pi_{p}\left(x_{\Pi}^{i}, y_{\Pi}^{i}, t_{\Pi}^{i}\right)\right|^{2}.\end{equation}

\noindent
where $N_b$ represents the number of training data relating input space to output space, and $N_{\Pi}$ represents the collocation points used to estimate the physical constraint term – see Fig \ref{fig:explain_inv_nn}. The $N_b$ data points can be sampled/measured from the whole domain, i.e., $\Omega \times \mathbb{T}$. All this data can contribute to both the $M S E_{b}$ and $M S E_{\Pi}$, i.e., $\{\cdot\}_{b}=\{\cdot\}_{\Pi}$ and $N_{b}=N_{\Pi}$, terms of the loss function. Also, we may include an extra set of $\{\cdot\}_{\Pi}$ (i.e., data where we would have no measured values), which would contribute only to the $M S E_{\Pi}$ term.

To minimize Eq. \ref{eq:loss_pinn_batch}, we train the neural network using the adaptive moment estimation (ADAM) algorithm \cite{kingma2014adam}. List \ref{list:batch_training} presents a code snippet used in this study. We will explore the performance as a function of the learning rate and batch size, while the other hyperparameters such as the number of hidden layers, $N_{h l}$, and number of neurons per layer, $N_n$, are fixed at 6 and 20, respectively, as these settings were
found to perform well for full batch training \cite{kadeethum2020pinn}. Besides, since there are many hyperparameters one could adjust in the training process, to see the effects of each specific parameter on training dynamics, we will use the one-factor-at-a-time method. The learning rate is a configurable hyper-parameters in the gradient descent method, and it is used to specify the amount that the weights (W and b) are updated during the backpropagation phase. If the learning rate is too high, the model becomes unstable, while if it is too small, the model takes a long time to converge. We use the following definition in our study \cite{keskar2016largebatch}:

\begin{equation} \text{batch} \ \text{size}=\left\{\begin{array}{cl}
1, & \text { SGD } \\
1<x<N_{t r}, & \text { batch } \\
N_{t r}, & \text { full }-\text { batch }
\end{array}\right.\end{equation}

\noindent
where SGD denotes Stochastic Gradient Descent. Note that we shuffle our training examples (validation and test sets are not altered) at every epoch. Note that one epoch is one complete presentation of the training data set to the learning algorithm. The neural networks are built on the Tensorflow platform \cite{tensorflow2015-whitepaper}. Besides, all training used in this study is performed on a single thread, Xeon Processor 2650v4.

%New colors defined below
\definecolor{codegreen}{rgb}{0,0.6,0}
\definecolor{codegray}{rgb}{0.5,0.5,0.5}
\definecolor{codepurple}{rgb}{0.58,0,0.82}
\definecolor{backcolour}{rgb}{0.95,0.95,0.92}

\lstdefinestyle{mystyle}{
  backgroundcolor=\color{backcolour},   commentstyle=\color{codegreen},
  keywordstyle=\color{magenta},
  numberstyle=\tiny\color{codegray},
  stringstyle=\color{codepurple},
  basicstyle=\ttfamily\footnotesize,
  breakatwhitespace=false,
  breaklines=true,
  captionpos=b,
  keepspaces=true,
  numbersep=5pt,
  showspaces=false,
  showstringspaces=false,
  showtabs=false,
  tabsize=2
}

\lstset{style=mystyle}
\begin{lstlisting}[language=Python, caption= Illustration of code snippet used in this study. The
learning rate and batch size are hyperparameters.]
import tensorflow as tf
import numpy as np

class PhysicsInformedNN:

    def __init__(self, x, y, t, u, v, p, layers, **kwargs):
        self.learning_rate = learning_rate
        self.optimizer_Adam = tf.train.AdamOptimizer(learning_rate=self.learning_rate)
        self.train_op_Adam = self.optimizer_Adam.minimize(self.loss_mean)

    def train_batch(self, nIter, batch_size):
        x_full, y_full, t_full, u_full, v_full, p_full \
            = self.x, self.y, self.t, self.u, self.v, self.p
        for i in range(nIter):
            x_full, y_full, t_full, u_full, v_full, p_full \
                = self.shuffle(x_full, y_full, t_full, u_full, v_full, p_full)
            for j in range( np.ceil(len(self.x) / batch_size).astype(int) ):
                idx_i = j*batch_size
                idx_e = (j+1)*batch_size

                tf_dict = {self.x_tf: self.x[idx_i:idx_e], self.y_tf: self.y[idx_i:idx_e],
                    self.t_tf: self.t[idx_i:idx_e],
                    self.u_tf: self.u[idx_i:idx_e], self.v_tf: self.v[idx_i:idx_e],
                    self.p_tf: self.p[idx_i:idx_e]}
                self.sess.run(self.train_op_Adam, tf_dict)

    def shuffle(self, x_full, y_full, t_full, u_full, v_full, p_full):
        s = np.arange(x_full.shape[0])
        np.random.shuffle(s)
        return x_full[s], y_full[s], t_full[s], u_full[s], v_full[s], p_full[s]
\end{lstlisting} \label{list:batch_training}

\section{Stochastic gradient descent, batch, or full-batch}

\begin{figure}[!ht]
   \centering
        \includegraphics[width=6.0cm, height=6.0cm]{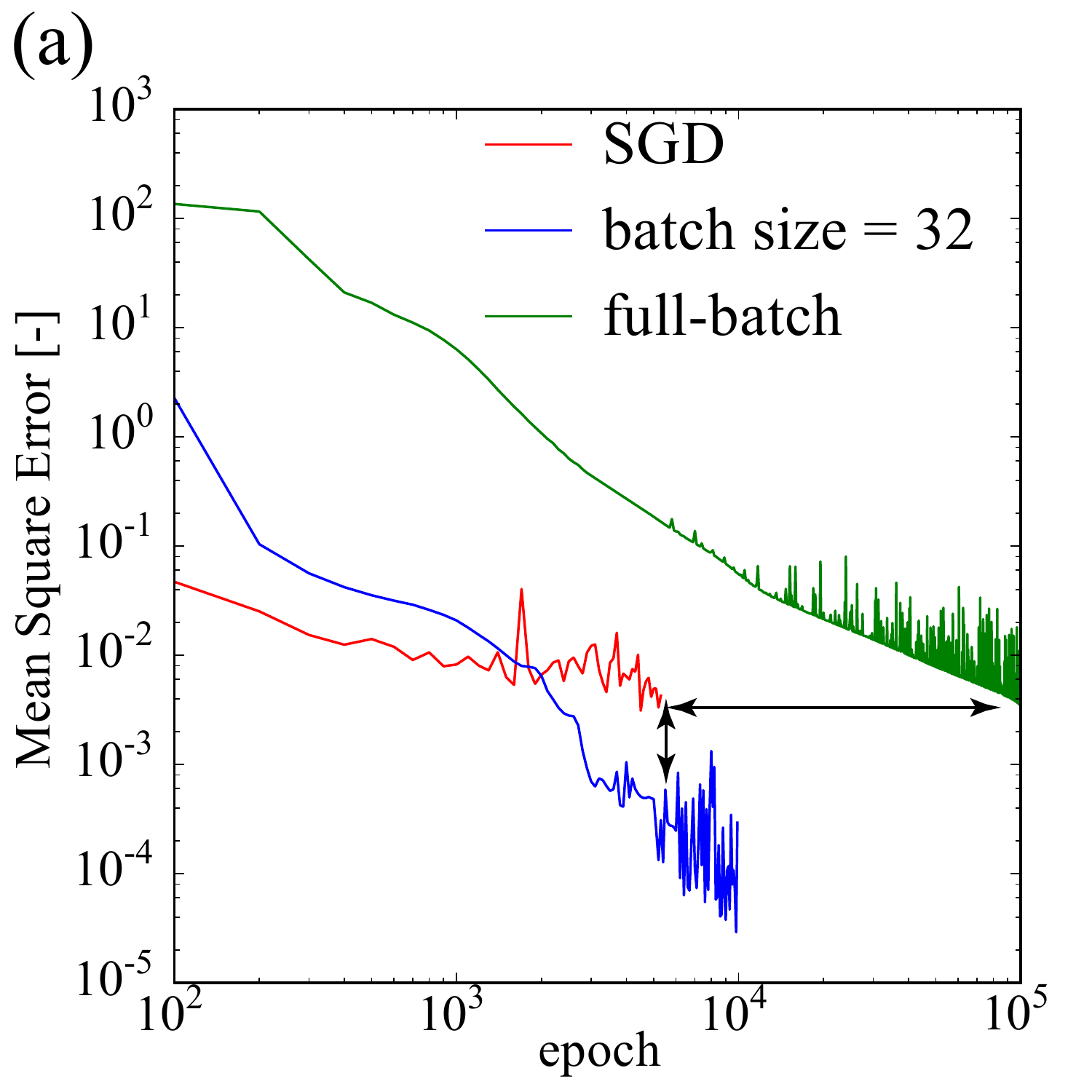}
        \includegraphics[width=6.0cm, height=6.0cm]{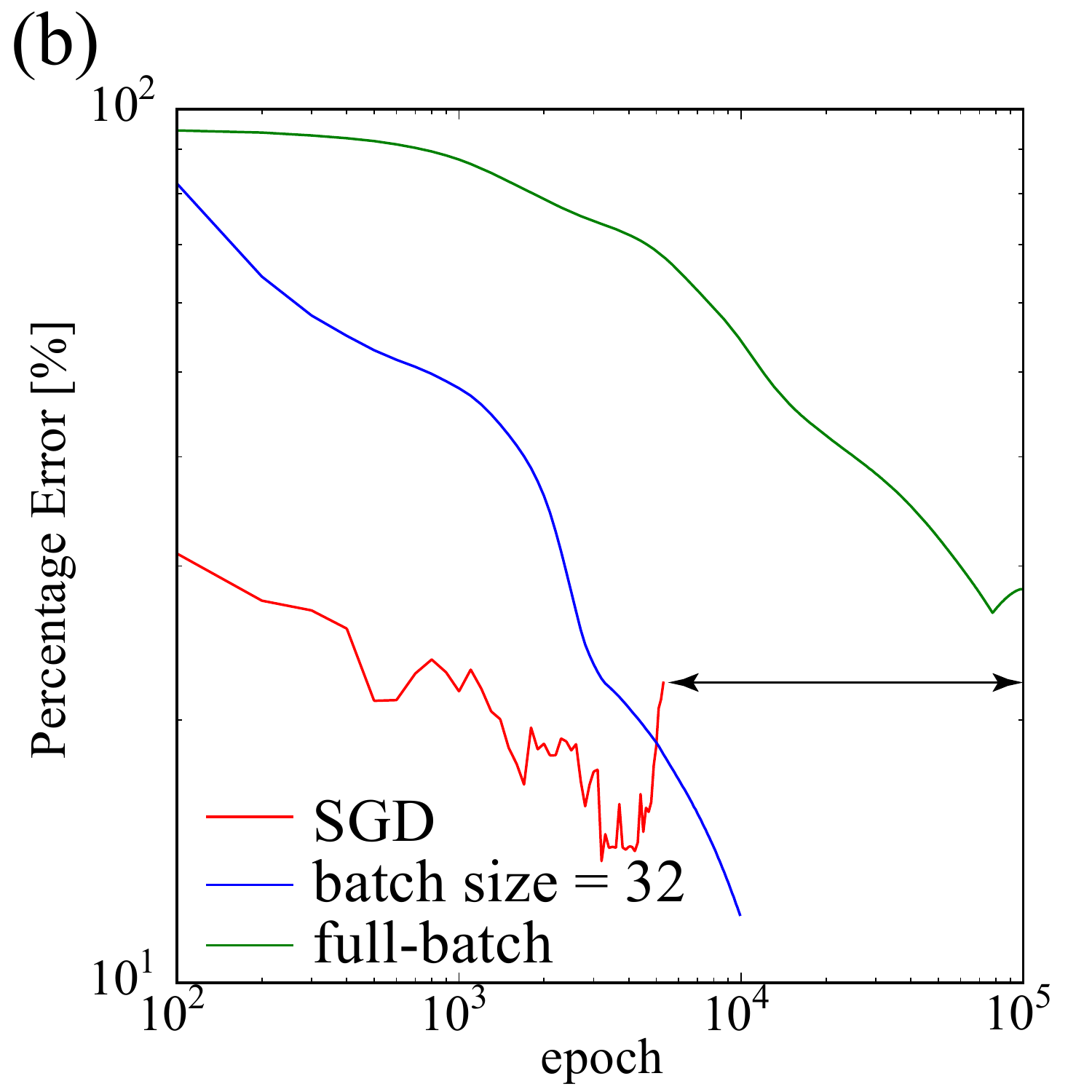}
   \caption{Comparison of stochastic gradient descent, batch, and
full-batch training effect w.r.t. (a) mean square errors and (b)
average percentage errors of $\theta$.}
   \label{fig:sgd_batch_full_batch}
\end{figure}

The comparison among SGD (batch size = 1), batch (batch size = 32), and full-batch training results is
presented in Fig \ref{fig:sgd_batch_full_batch}. The learning rate is fixed at 0.0005. Fig  \ref{fig:sgd_batch_full_batch}a shows that the $MSE$ of the SGD method decays slowly and fluctuates when epoch $> 1000$. The $MSE$ of the full-batch training needs $10^5$ epochs to reach the level the SGD obtains after only 5000 epochs. The batch training seems to have the best performance among these three methods. Fig \ref{fig:sgd_batch_full_batch}b shows the average percentage errors of $\theta$, confirming that batch training performs better as a function of the number of epochs.

Next, we compare the wall time measured at epoch = 5000, and the results are shown in Table \ref{tab:1}. The SGD training has a much higher wall time than the batch and full-batch training. Note that wall time or elapsed time is measured from the start to the end of each process. Moreover, the full-batch training requires the least wall time. We observe that the batch training results in the best accuracy for both $u$, $v$, and $p$ and $\theta$ while it, however, requires a reasonable wall time. In the following, we will explore the batch training in more detail.

\begin{table}[htbp]
  \centering
  \caption{Wall time comparison among SGD, batch, and full-
batch model training (Hardware: Xeon Processor 2650v4)}
    \begin{tabular}{|c|c|c|}
    \hline
    Method & avg. time/100 epochs & total time (5000 epochs) \\
    \hline
    \hline
    SGD   & 5876 s. & 299699 s. \\
    \hline
    batch & 262 s. & 13139 s. \\
    \hline
    full-batch & 69 s. & 3479 s. \\
    \hline
    \end{tabular}%
  \label{tab:1}%
\end{table}%

\section{Effect of batch size on training dynamics}

\begin{figure}[!ht]
   \centering
        \includegraphics[width=6.0cm, height=6.0cm]{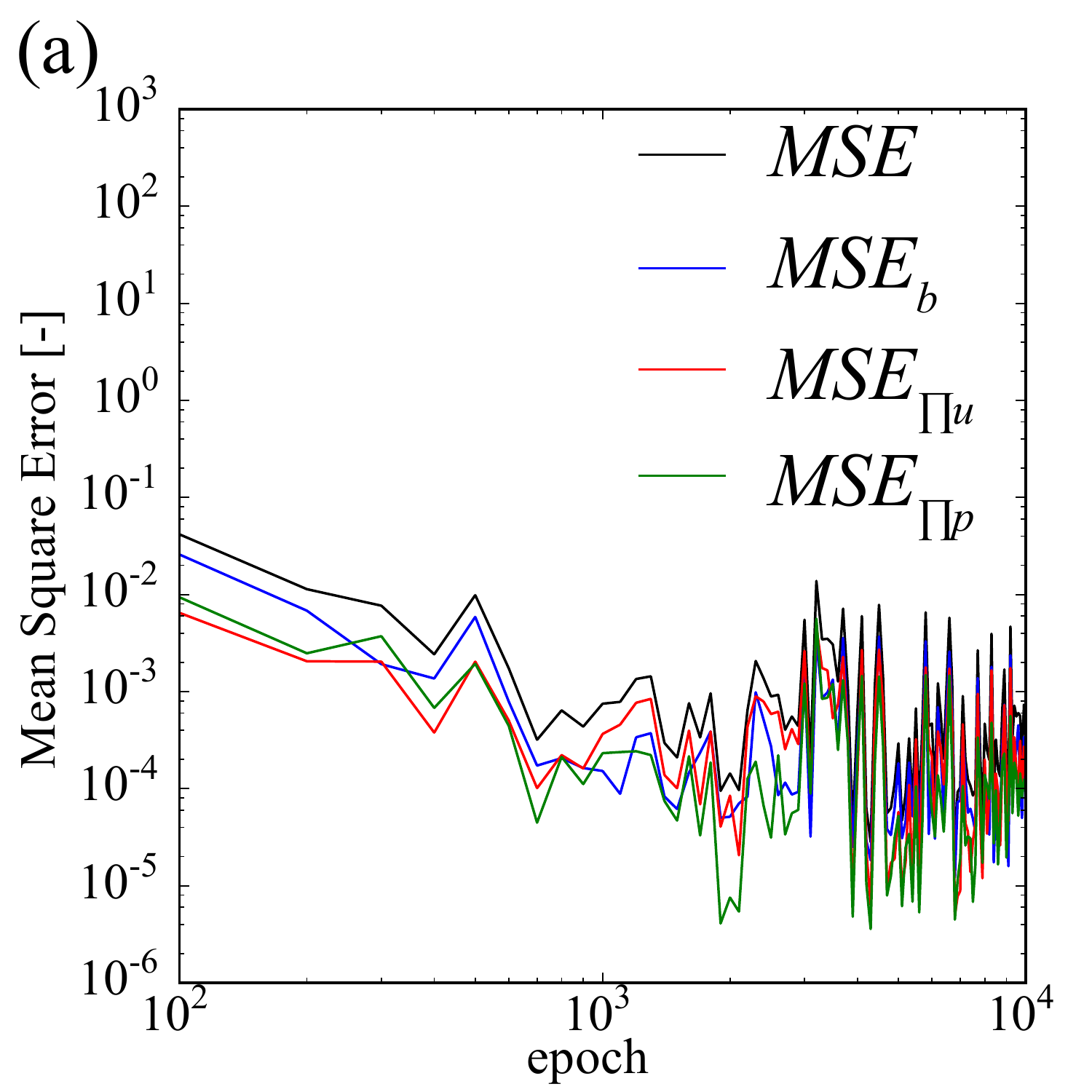}
        \includegraphics[width=6.0cm, height=6.0cm]{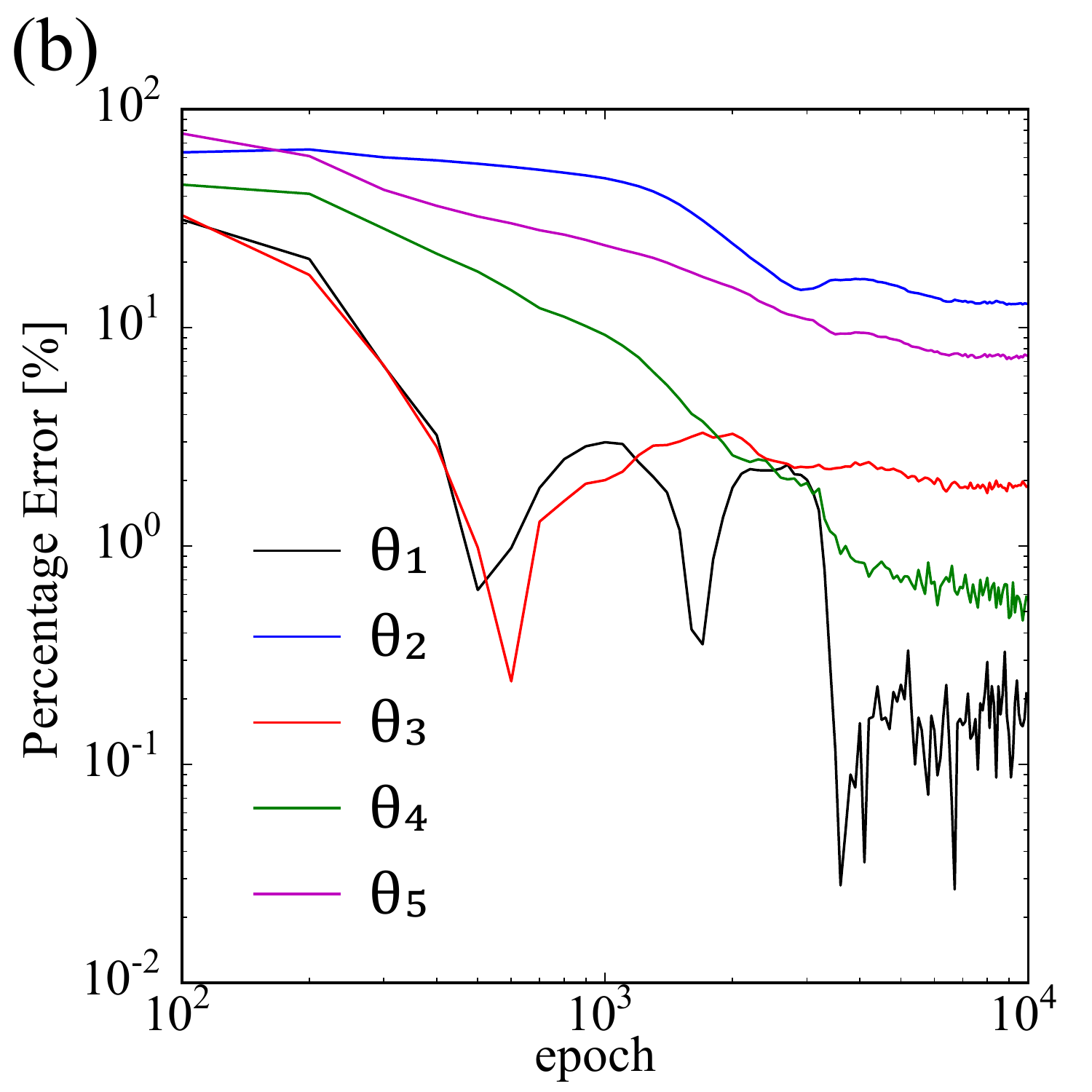}
        \includegraphics[width=6.0cm, height=6.0cm]{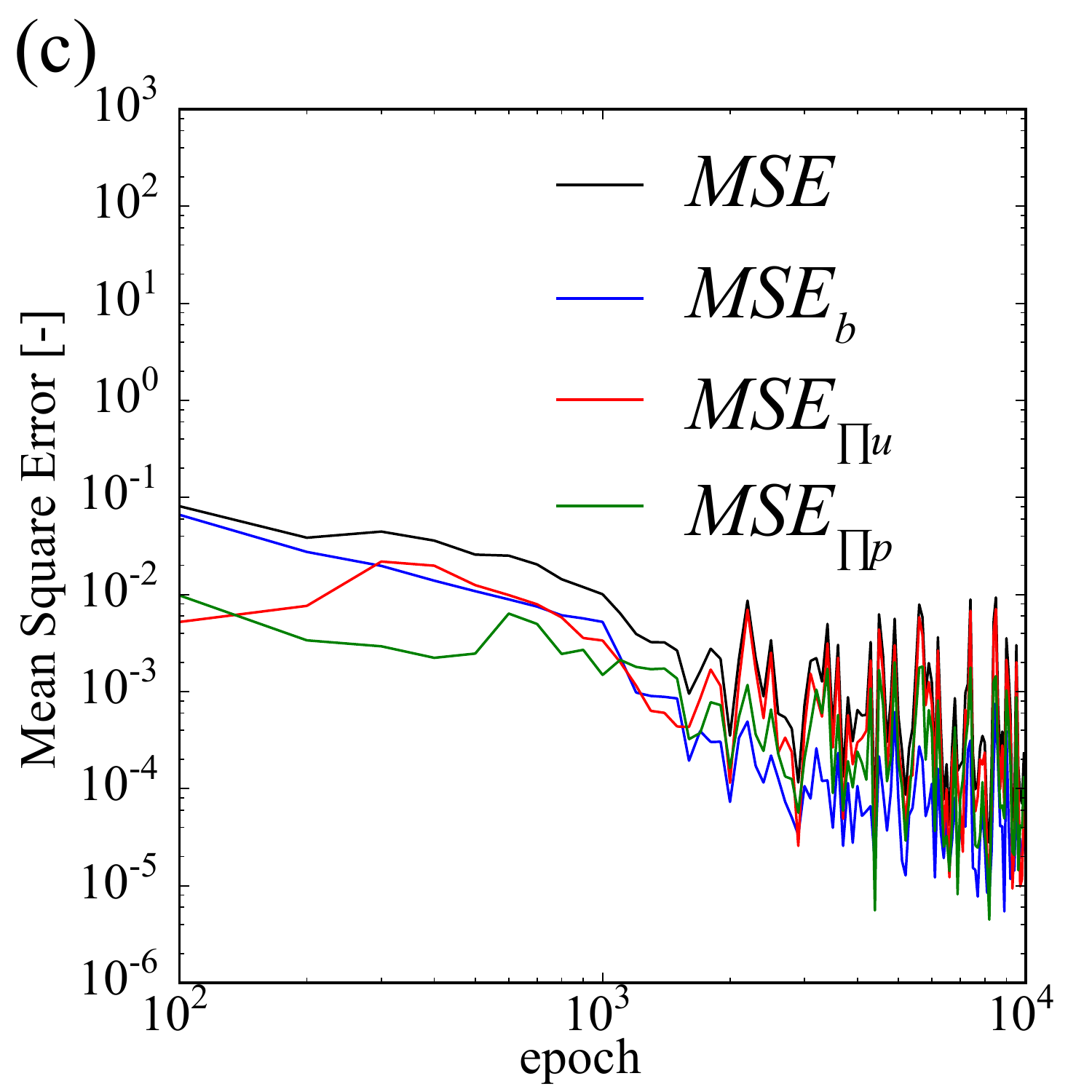}
        \includegraphics[width=6.0cm, height=6.0cm]{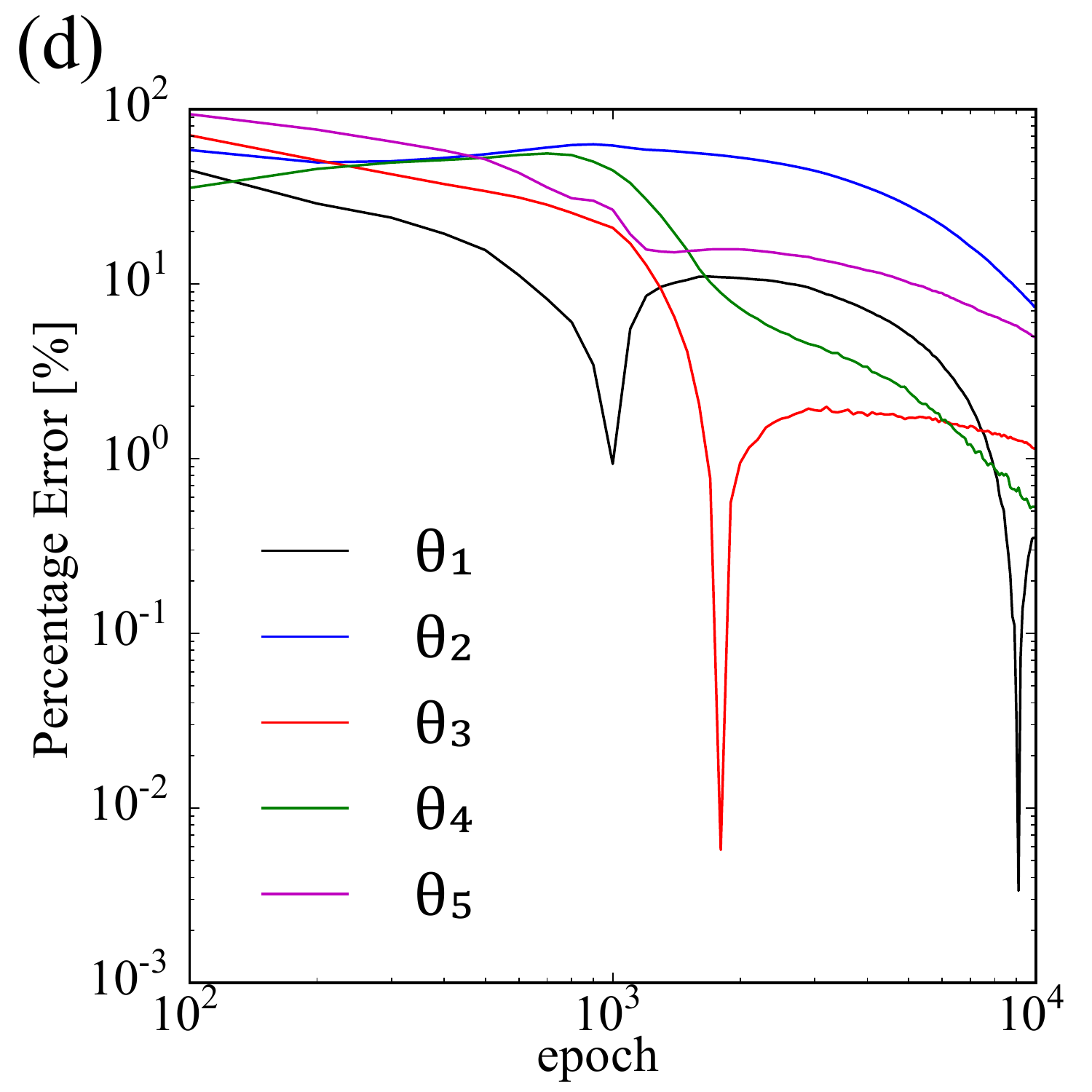}
        \includegraphics[width=6.0cm, height=6.0cm]{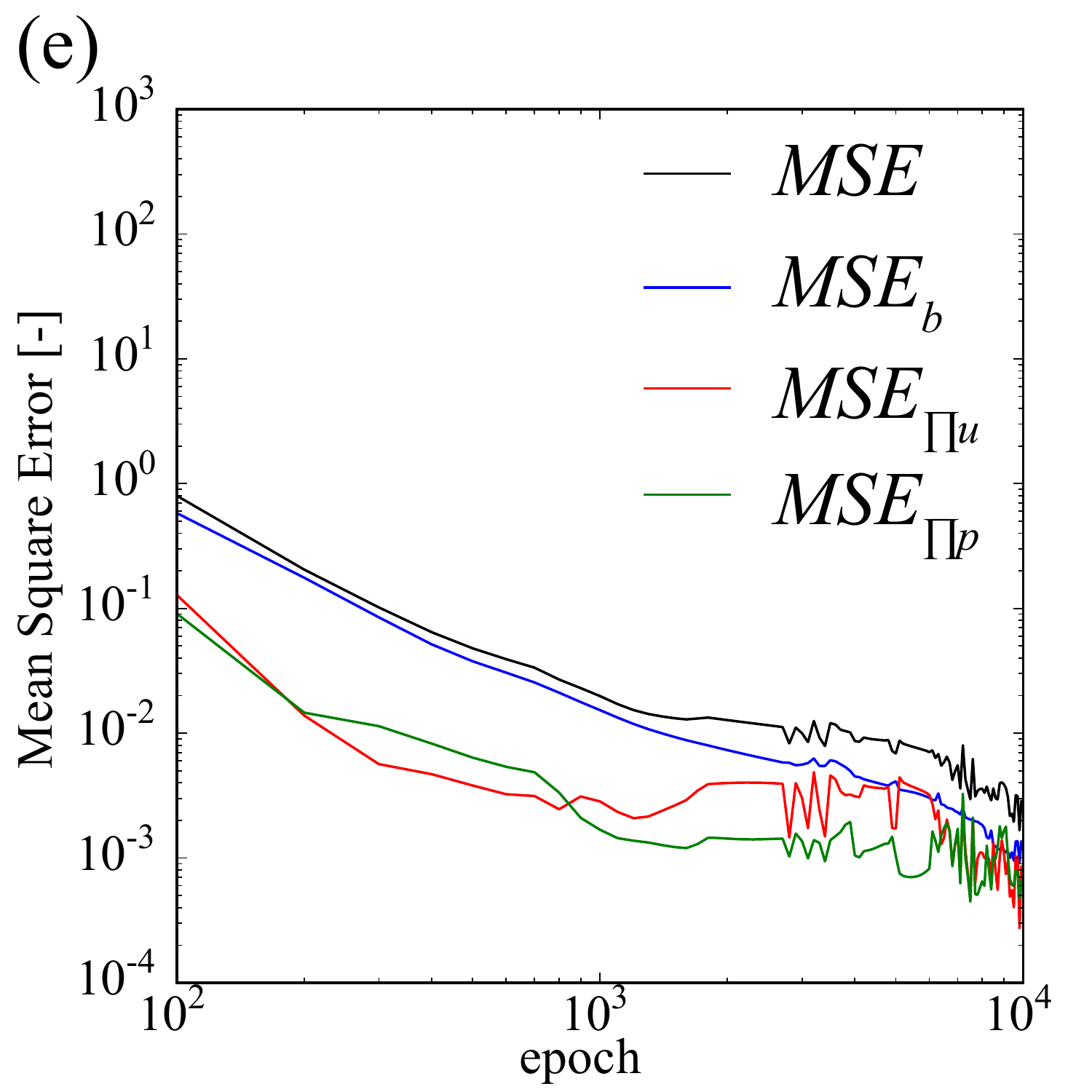}
        \includegraphics[width=6.0cm, height=6.0cm]{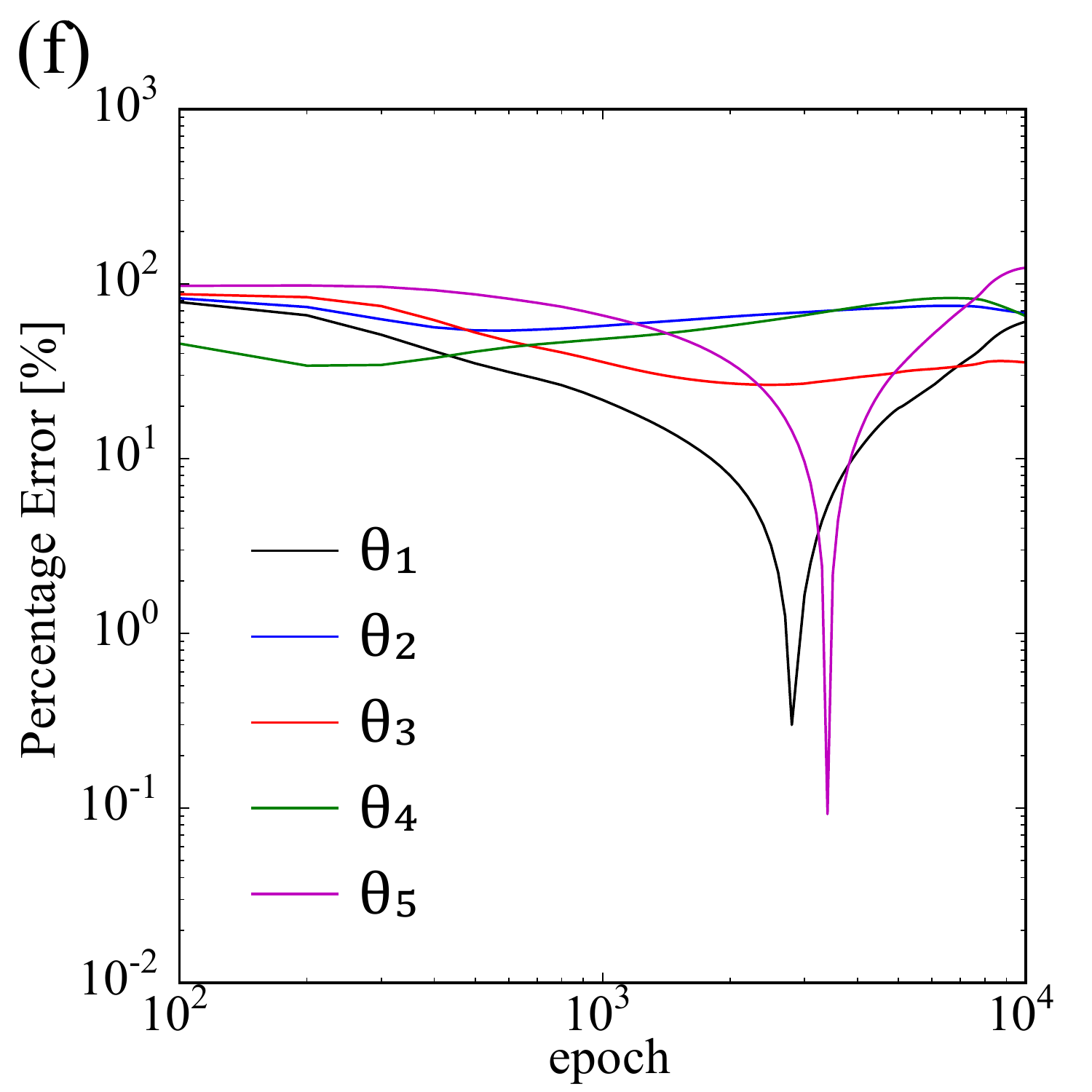}
   \caption{Illustration of the effect of batch size on training
dynamics: (a) batch size of 8 – mean square errors of the loss
functions, (b) – percentage errors of $\theta$, (c) batch size of 32 –
mean square errors of the loss functions, (d) – percentage
errors of $\theta$, (e) batch size of 128 – mean square errors of the
loss functions, (f) – percentage errors of $\theta$. The learning rate is
fixed at 0.0005.}
   \label{fig:4}
\end{figure}

The effect of batch size on training dynamics is presented in Fig \ref{fig:4}. Here, we use batch size = 8, 32, and 128. Note that the learning rate is fixed at 0.0005 for this study. From Fig  \ref{fig:4}, one can observe that with the larger batch sizes, both the dynamics of the mean square errors and the percentage errors of $\theta$ become smoother (less fluctuations). With a fixed number of the epoch, the batch size of 128 has the worst accuracy, while the batch size of 8 produces the largest fluctuations. Using the batch size of 32, we here obtain the best performance with respect to minimizing the loss functions.

\begin{figure}[!ht]
   \centering
        \includegraphics[width=6.0cm, height=6.0cm]{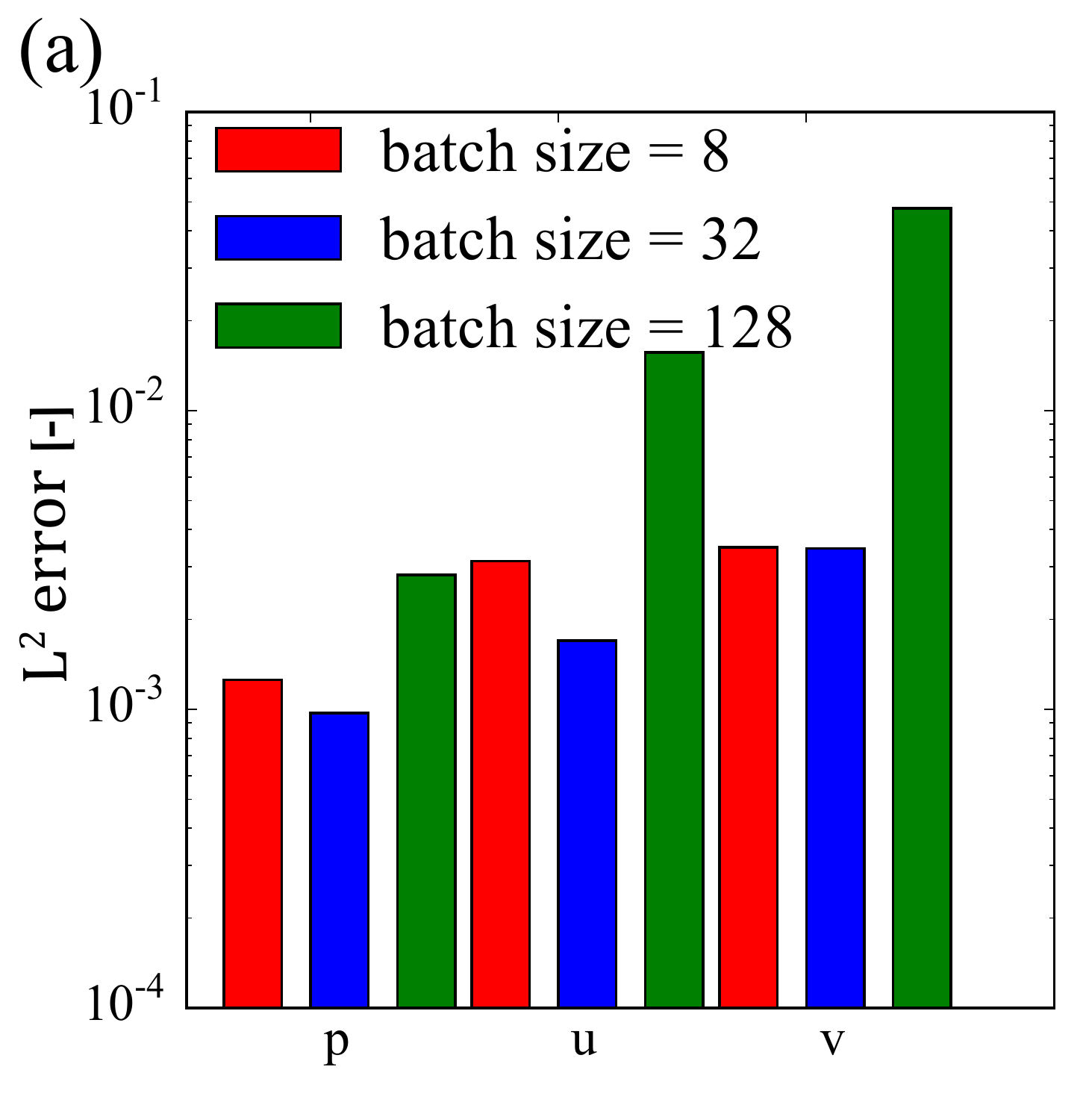}
        \includegraphics[width=6.0cm, height=6.0cm]{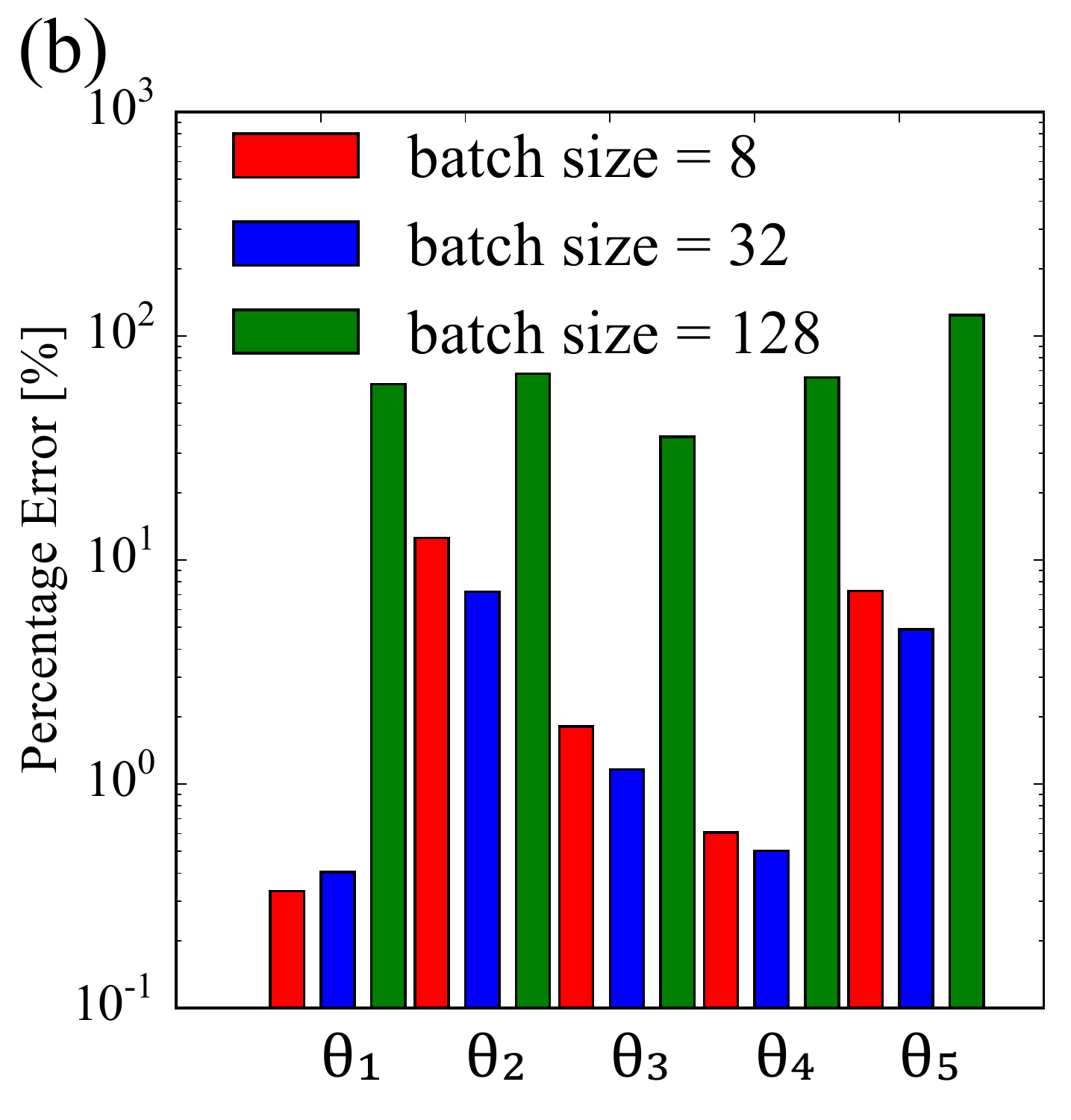}
   \caption{Illustration of the accuracy of the trained model tested
using the validation set ($u$, $v$, and $p$) and the true values of the
physical parameters ($\theta_{1}, \theta_{2}, \theta_{3}, \theta_{4},$ and $\theta_{5}$): (a) L2 errors of $u$, $v$, and $p$ and (b) percentage errors of $\theta_{1}, \theta_{2}, \theta_{3}, \theta_{4},$ and $\theta_{5}$.}
   \label{fig:5}
\end{figure}

The comparison of the accuracy of the trained model, when tested on the validation set as a function of the batch size, is shown in Fig \ref{fig:5}. The results illustrate that using the batch size of 32 generally produces the most accurate results, while the trained model with the batch size of 128 has the least accurate results.

The time comparison among three batch sizes is provided in Table \ref{tab:2}. As expected, the larger batch sizes require less training time. Note that the larger batch size requires less backward propagation and weight updating operations; therefore, it takes less time to complete one epoch.

\begin{table}[htbp]
  \centering
  \caption{Wall time comparison among the batch size of 8, 32,
and 128 model training (Hardware: Xeon Processor 2650v4)}
    \begin{tabular}{|c|c|c|}
    \hline
    {batch} & avg. time/100 epochs & total time (10000 epochs) \\
    \hline
    \hline
    8     & 1332 s. & 133253 s. \\
    \hline
    32    & 262 s. & 26269 s. \\
    \hline
    128   & 146 s. & 14646 s. \\
    \hline
    \end{tabular}%
  \label{tab:2}%
\end{table}%

\section{Effect of learning rate on training dynamics}

\begin{figure}[!ht]
   \centering
        \includegraphics[width=6.0cm, height=6.0cm]{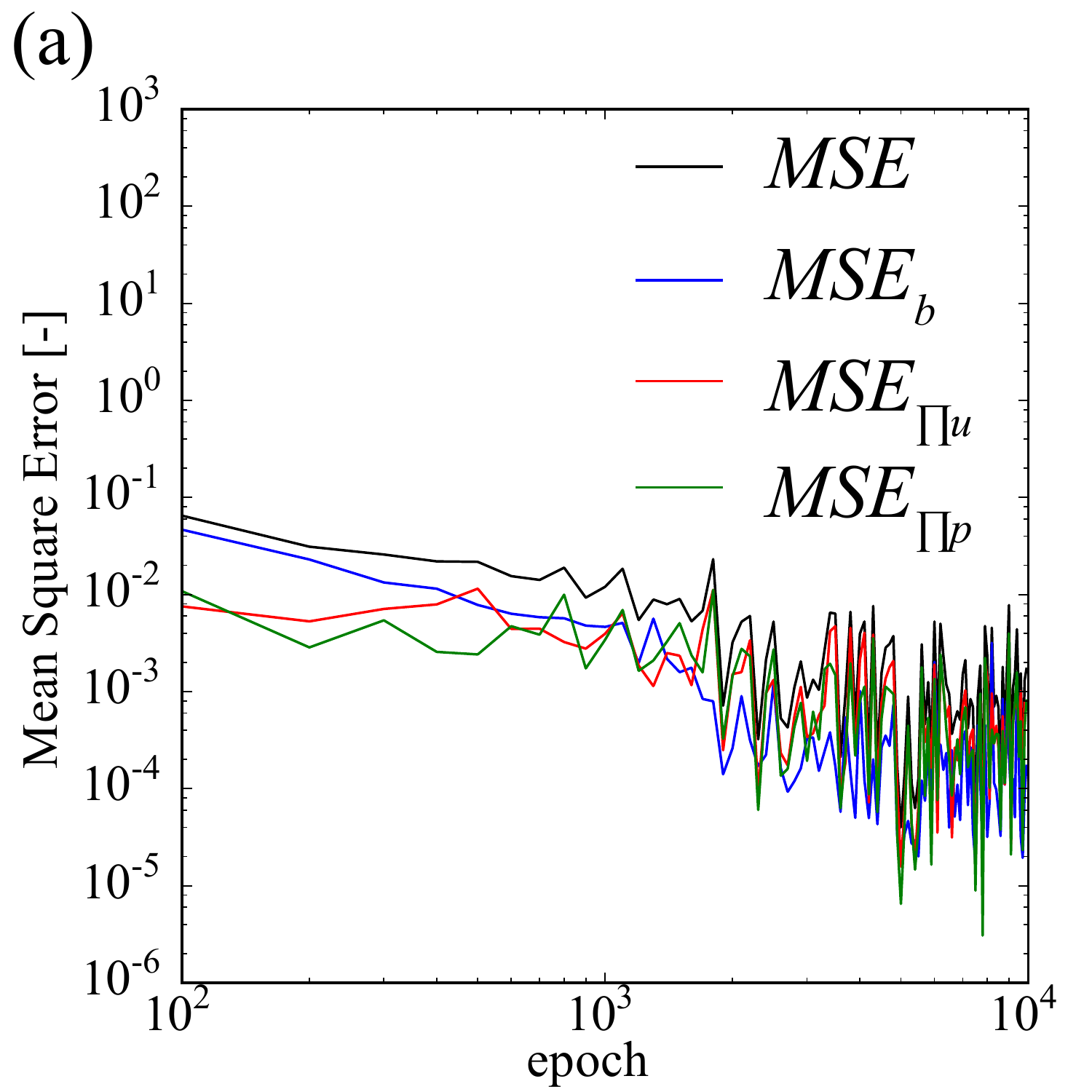}
        \includegraphics[width=6.0cm, height=6.0cm]{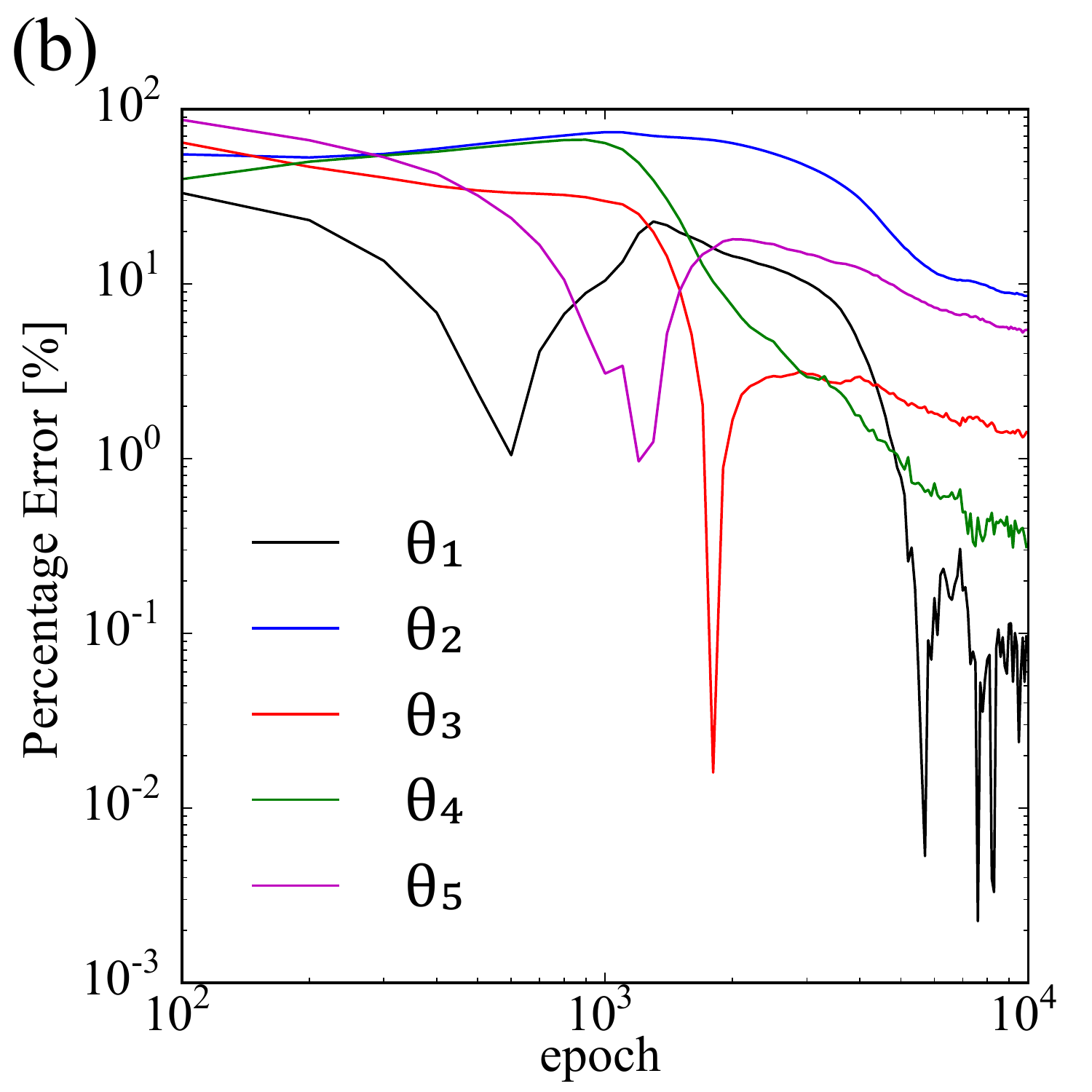}
        \includegraphics[width=6.0cm, height=6.0cm]{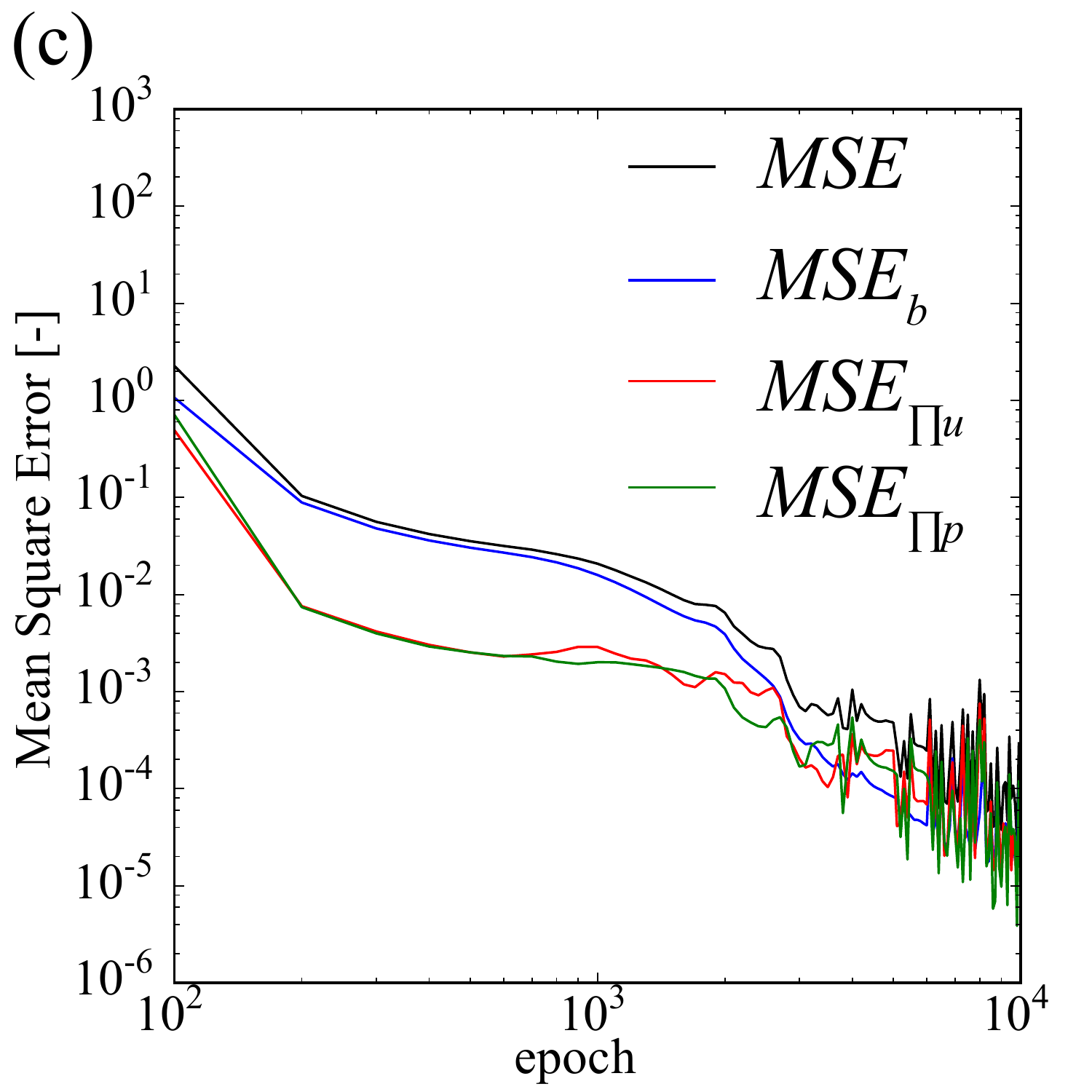}
        \includegraphics[width=6.0cm, height=6.0cm]{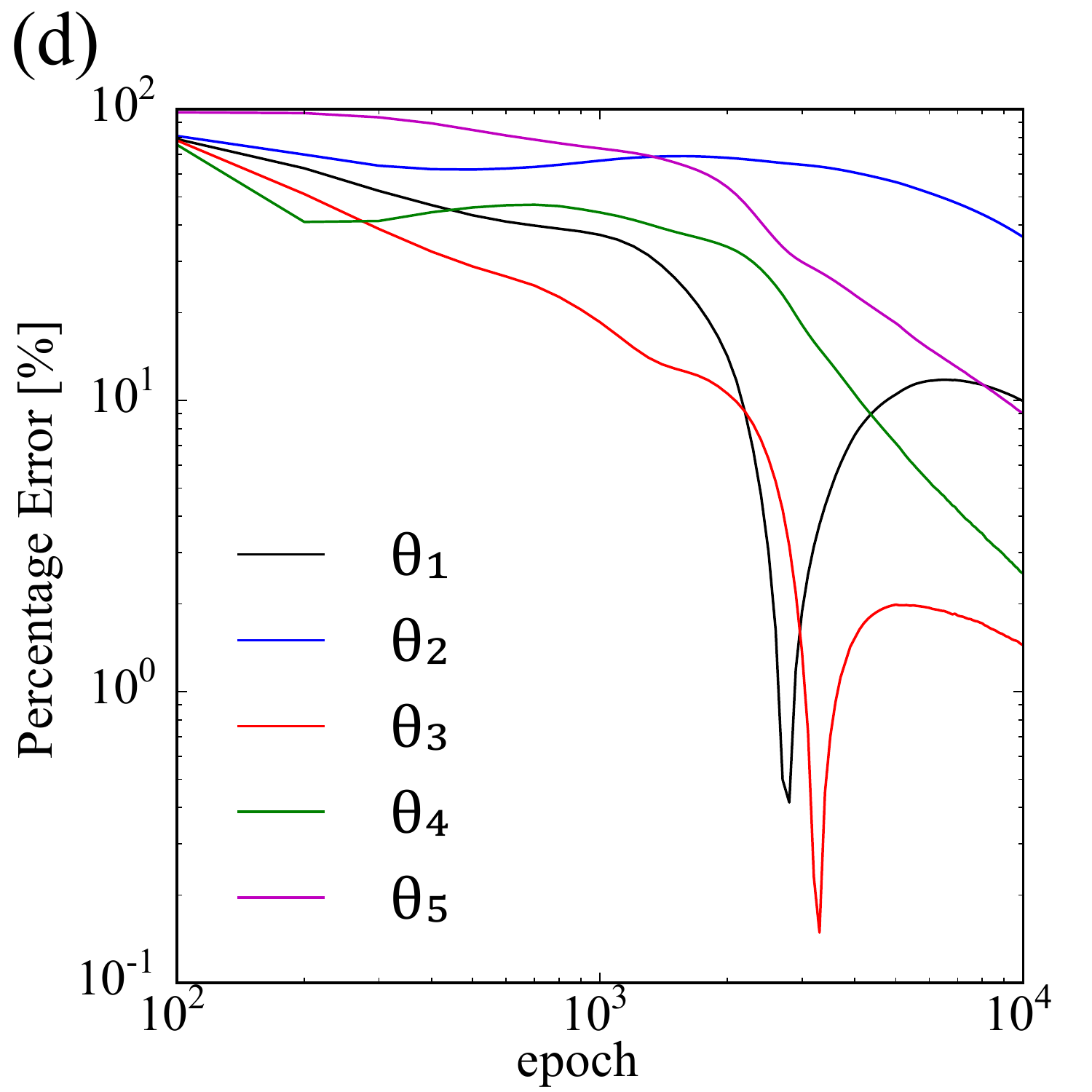}
   \caption{Illustration of effect of learning rate on training
dynamics: (a) learning rate of 0.001 – mean square errors of
the loss functions, (b) – percentage errors of $\theta$, (c) learning
rate of 0.0001 – mean square errors of the loss functions, (d) –
percentage errors of $\theta$. The batch size is fixed at 32. Please
refer to Fig \ref{fig:4}c-d for the results of the learning rate of 0.0005.}
   \label{fig:6}
\end{figure}

Next, we investigate the effect of the learning rate on training behavior. Here, we examine three different learning rates, 0.001, 0.0005, and 0.0001, respectively. The results are shown in Fig \ref{fig:6}. Note that the batch size is fixed at 32 in this investigation. As expected, when the learning rate increases, the oscillation in both mean square errors and percentage errors of $\theta$ become more substantial.

\begin{figure}[!ht]
   \centering
        \includegraphics[width=6.0cm, height=6.0cm]{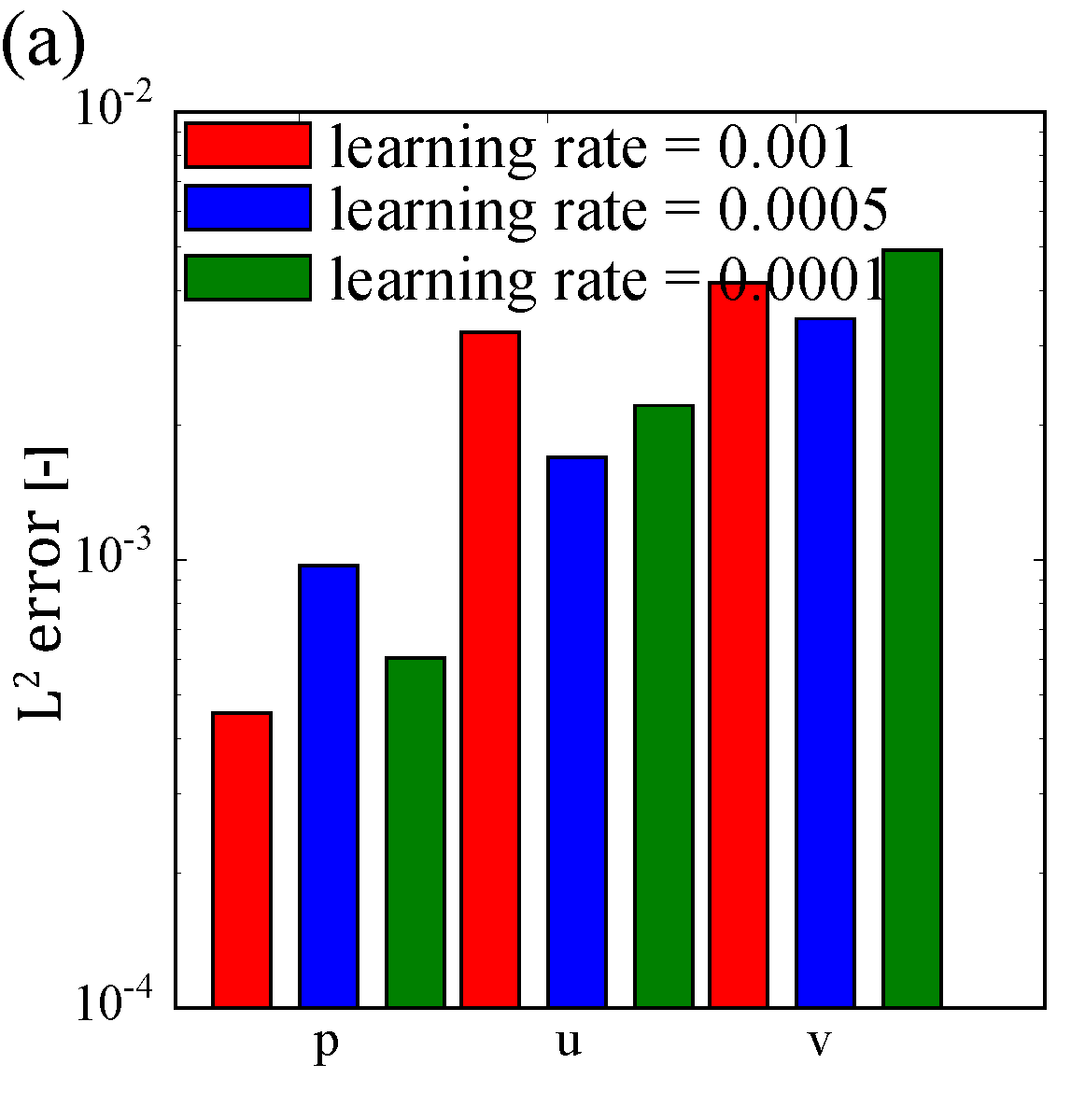}
        \includegraphics[width=6.0cm, height=6.0cm]{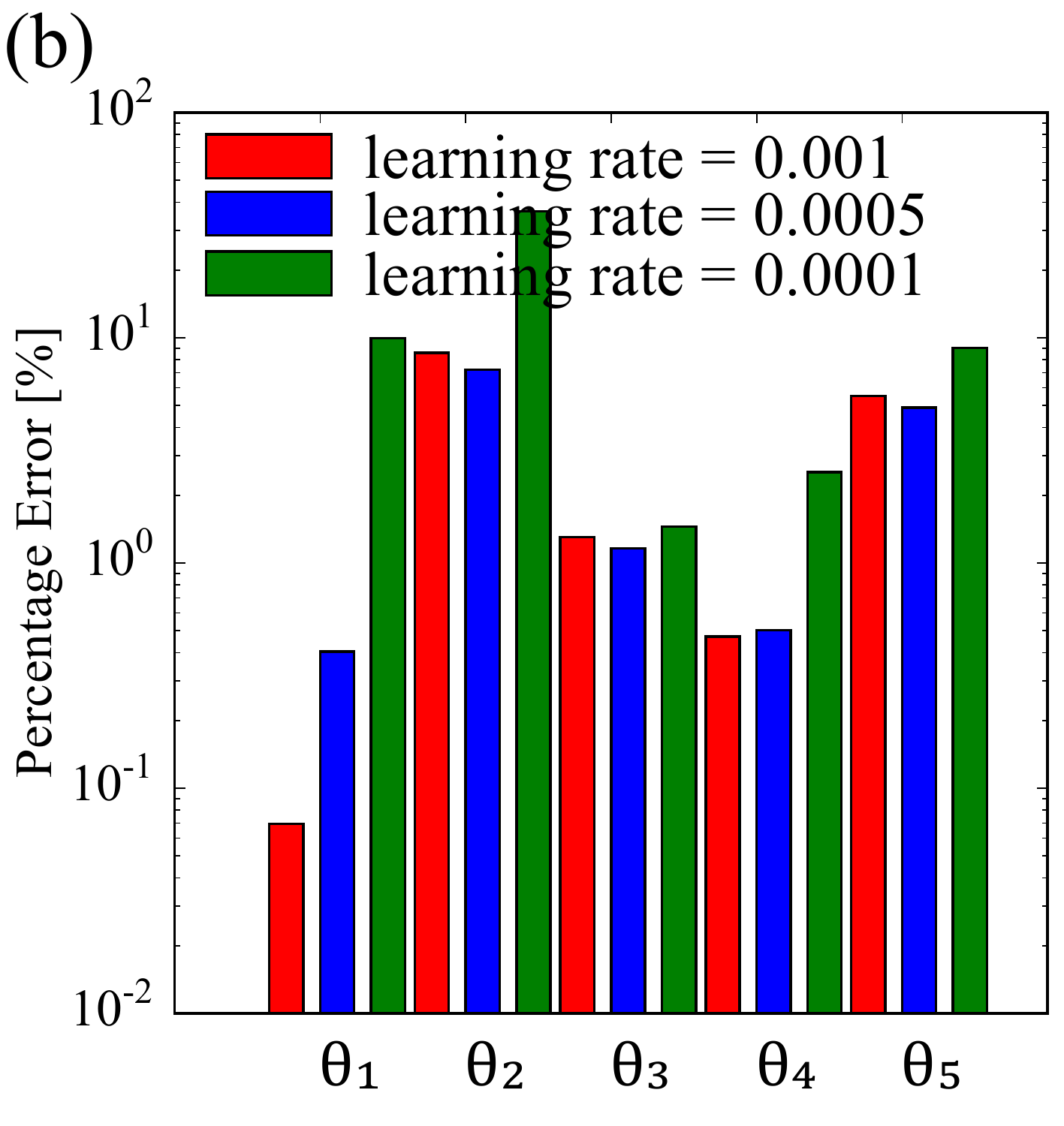}
   \caption{Illustration of the accuracy of the trained model tested
using the validation set ($u$, $v$, and $p$) and the true values of the
physical parameters ($\theta_{1}, \theta_{2}, \theta_{3}, \theta_{4},$ and $\theta_{5}$): (a) L2 errors of $u$, $v$, and $p$ and (b) percentage errors of $\theta_{1}, \theta_{2}, \theta_{3}, \theta_{4},$ and $\theta_{5}$. Please
refer to Fig \ref{fig:4}c-d for the results of the learning rate of 0.0005.}
   \label{fig:7}
\end{figure}

The accuracy of the trained model, when tested on the validation set, is presented in Fig \ref{fig:7}. Generally, the trained model using the learning rate of 0.0005 yields the most accurate results.

\section{Effect of weights and biases initialization on training dynamics}

\begin{figure}[!ht]
   \centering
        \includegraphics[width=6.0cm, height=6.0cm]{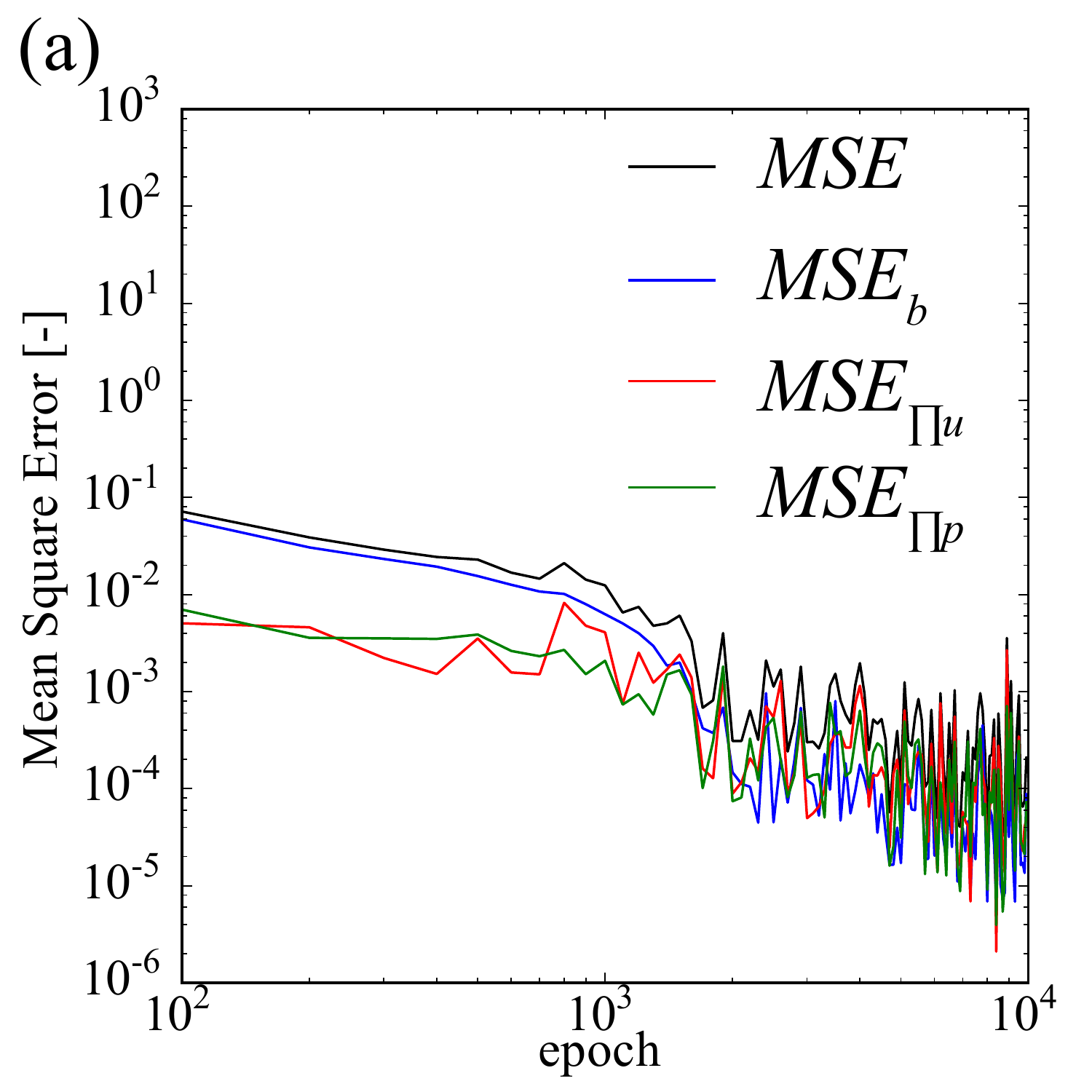}
        \includegraphics[width=6.0cm, height=6.0cm]{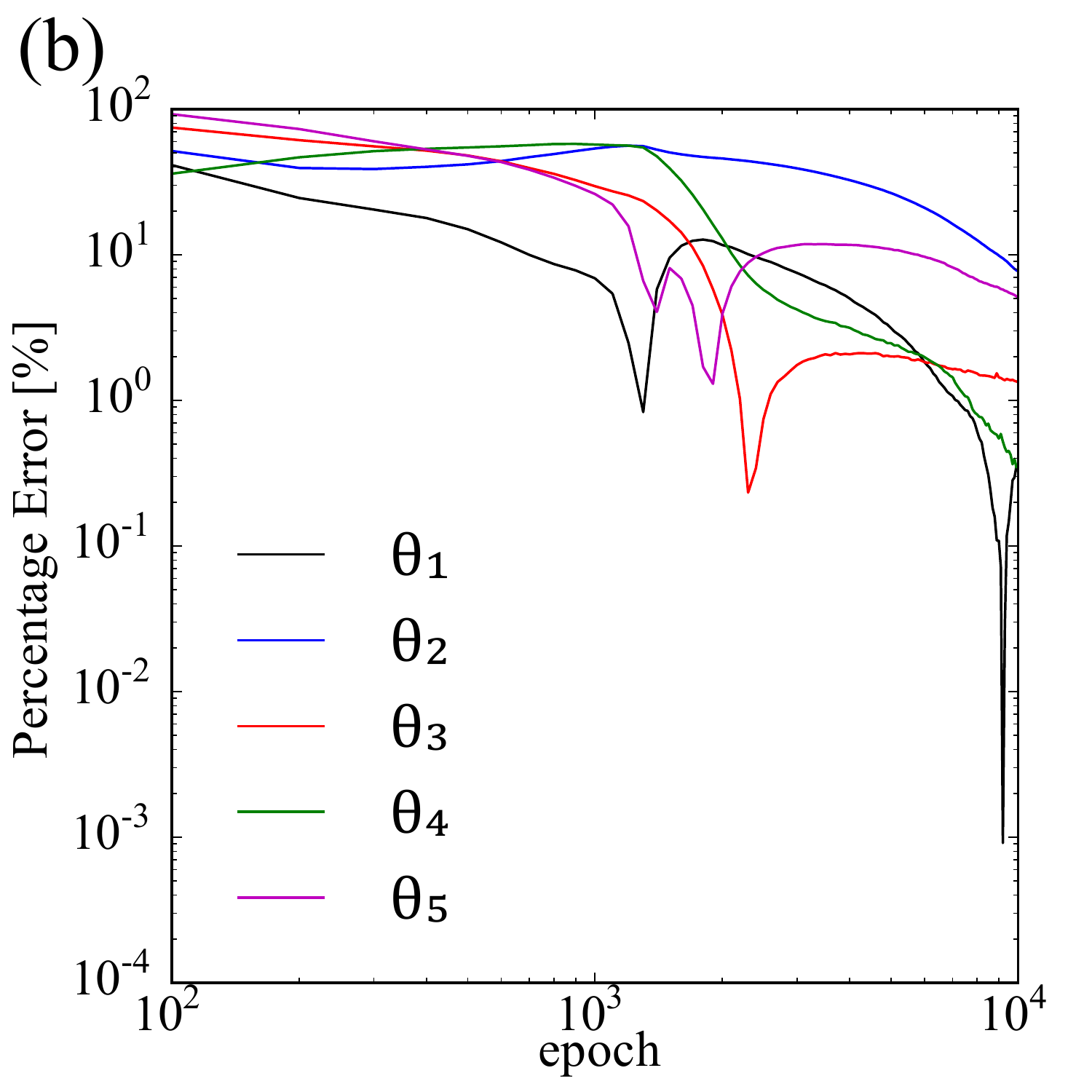}
        \includegraphics[width=6.0cm, height=6.0cm]{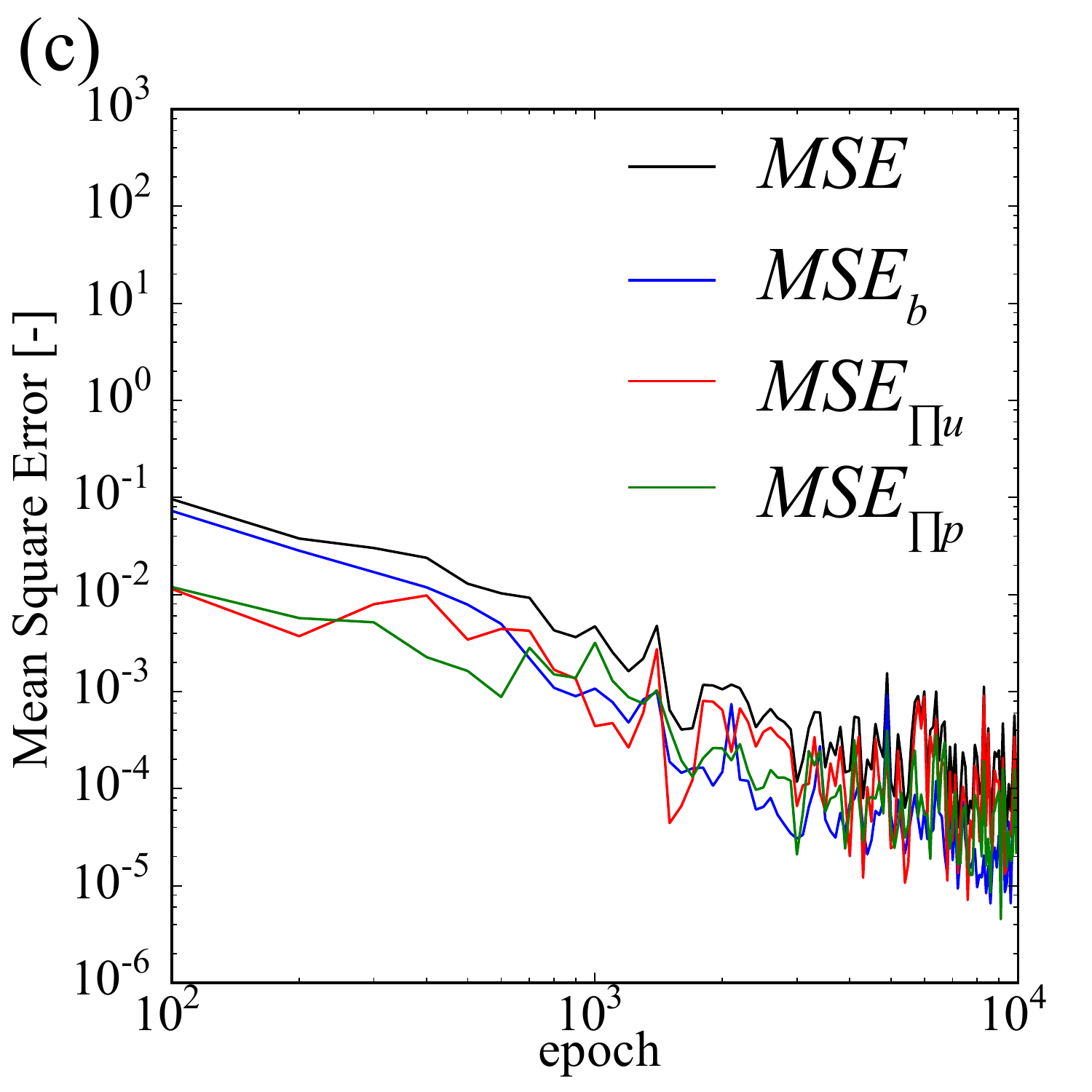}
        \includegraphics[width=6.0cm, height=6.0cm]{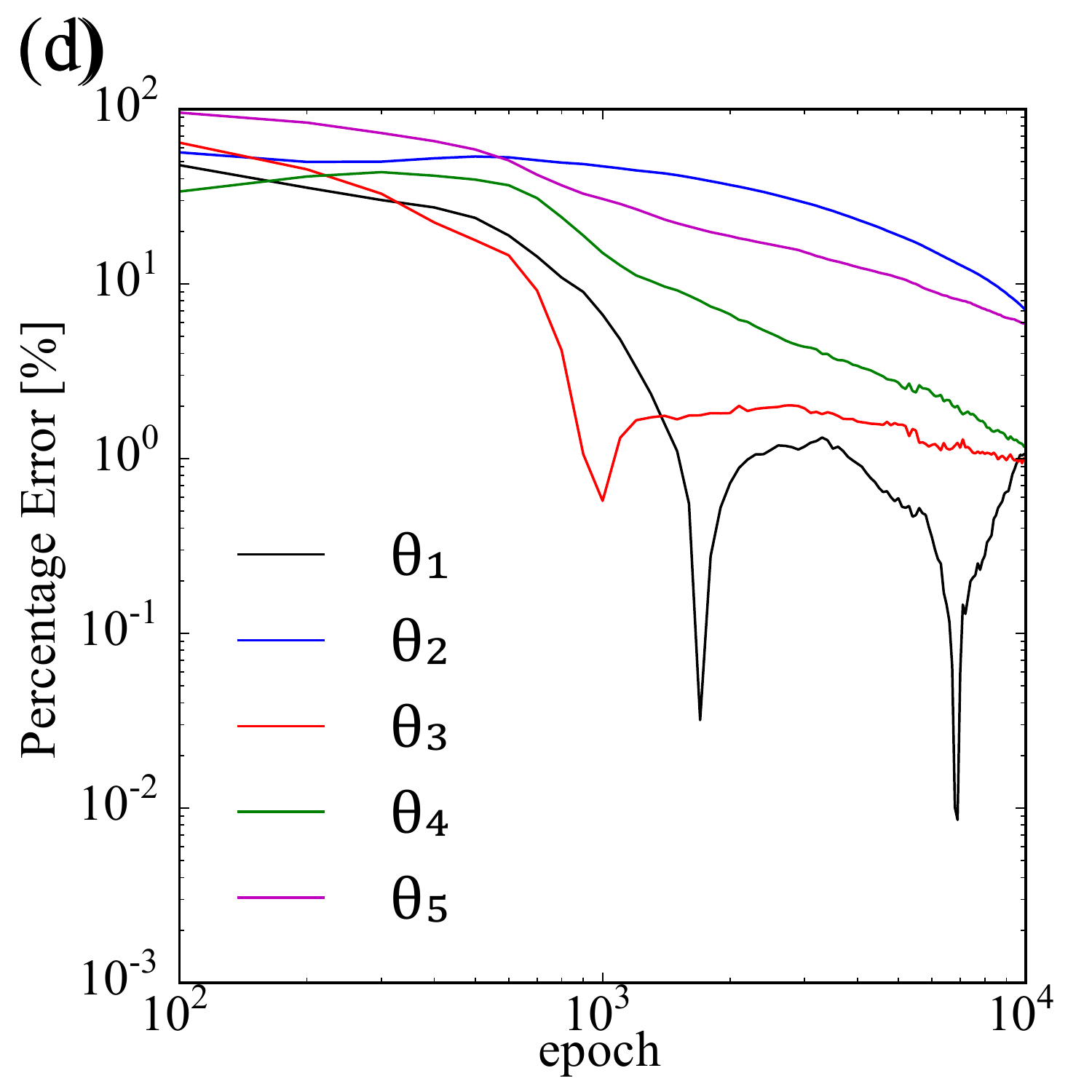}
   \caption{Illustration of the effect of weights and biases
initialization on training dynamics: (a) init1 – mean square
errors of the loss functions, (b) – percentage errors of $\theta$, (c)
init3 – mean square errors of the loss functions, (d) –
percentage errors of $\theta$. The batch size is fixed at 32, and the
learning rate is fixed at 0.0005. Please refer to Fig \ref{fig:4}c-d for the
results of init2.}
   \label{fig:8}
\end{figure}

In the previous sections, we found that the trained model with batch size = 32 and learning rate = 0.0005 generally provides the most accurate predictions of $u$, $v$, $p$, and the estimations of the set $\theta$. However, as mentioned in \cite{kadeethum2020pinn}, PINN training depends heavily
on the initializations of W and b. Hence, we present the effects of W and b initializations in Fig \ref{fig:8}. Note that we use “Xavier initialization” for all of our simulations. Since Xavier initialization is stochastic, the sets of initial W and b are different with different seed numbers. The goal here
is to illustrate that with different initial sets of W and b, the training behavior of each initialization is different, and, as mentioned in \cite{kadeethum2020pinn}, one could
report an average of the results instead of one single outcome. Here, we use batch size = 32 and learning rate = 0.0005. As expected, the convergent paths are different among different initializations.

\begin{figure}[!ht]
   \centering
        \includegraphics[width=6.0cm, height=6.0cm]{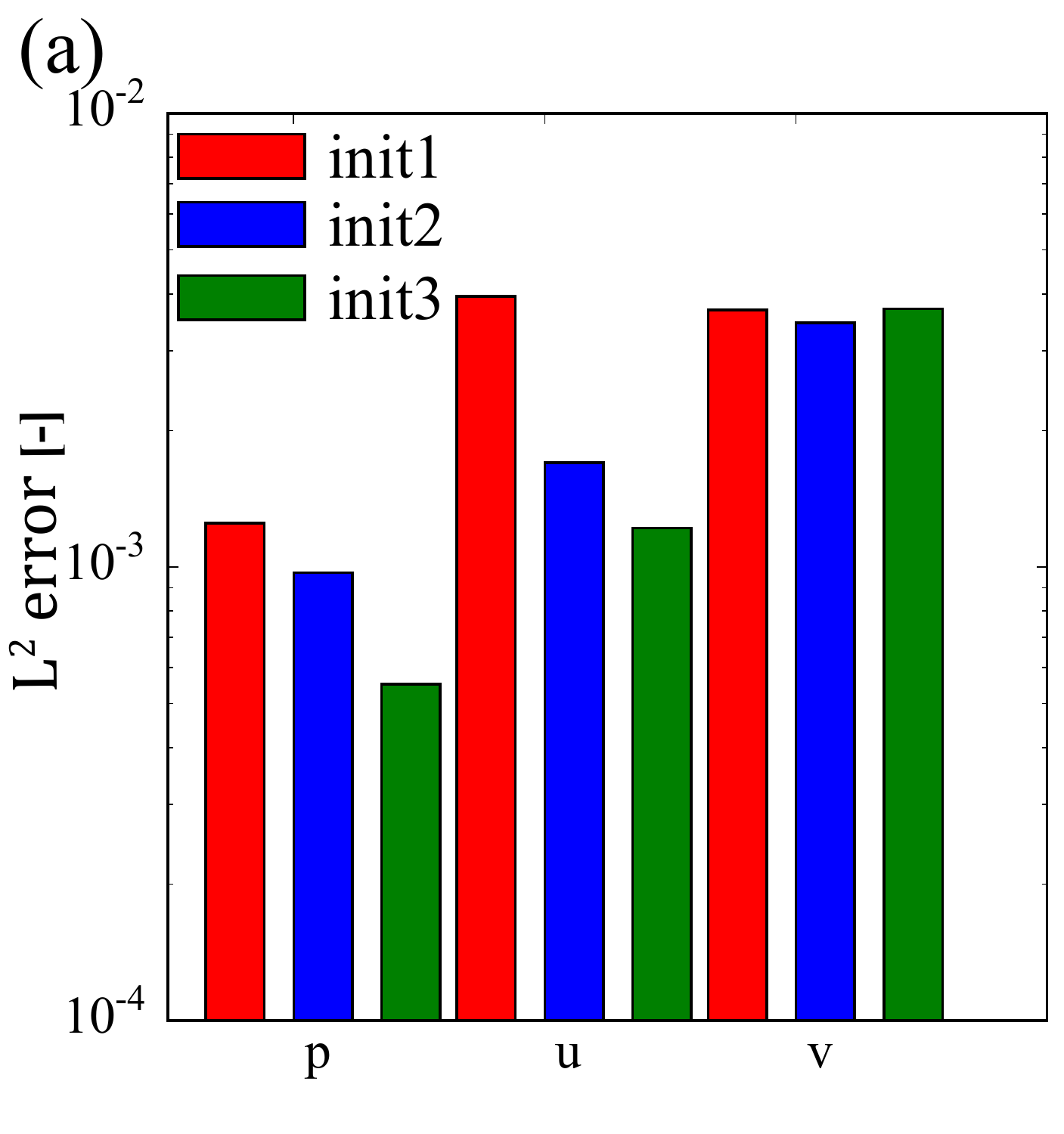}
        \includegraphics[width=6.0cm, height=6.0cm]{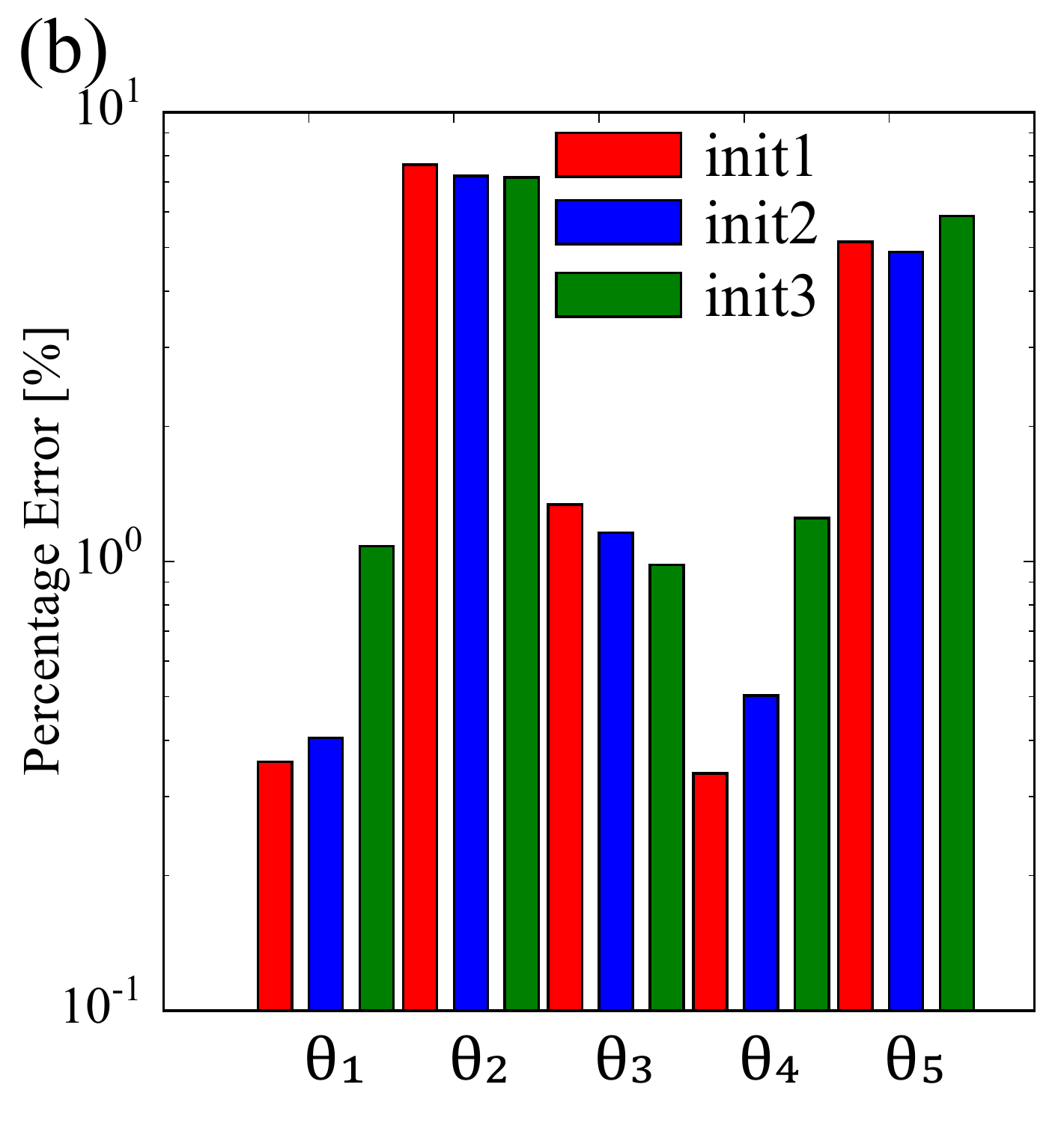}
   \caption{Illustration of the accuracy of the trained model tested
using the validation set ($u$, $v$, and $p$) and the true values of the
physical parameters ($\theta_{1}, \theta_{2}, \theta_{3}, \theta_{4},$ and $\theta_{5}$): (a) L2 errors of $u$, $v$, and $p$ and (b) percentage errors of $\theta_{1}, \theta_{2}, \theta_{3}, \theta_{4},$ and $\theta_{5}$. Please refer to Fig \ref{fig:4}c-d for the
results of init2.}
   \label{fig:9}
\end{figure}

Even though the convergent paths among different
initializations are dissimilar, the accuracy of the trained
model tested against the validation set for $u$, $v$, and $p$ and
true values for a set of $\theta$ is not much different, as shown
in Fig \ref{fig:9}.

\section{Effect of noise on model behavior}

Next, we perform a systematic study of the effect of
additive noise in data, which is created from the true data
as follows \cite{raissi2019physics}:

\begin{equation}\label{eq:darcy_noise}
\bm{X}_{noise} = \bm{X}_{true} + \epsilon \mathcal{S}\left(\bm{X}_{true}\right) \mathcal{G}\left(0,1\right),
\end{equation}

\noindent
where $\bm{X}_{noise}$ and $\bm{X}_{true}$ is the vector of the data with and without noise, respectively. The $\epsilon$ determines the noise level, $\mathcal{S}\left(\cdot\right)$ represents a standard deviation operator, $\mathcal{G}\left(0,1\right)$ is a random value, which is sampled from the Gaussian distribution with mean and standard deviation of zero and one, respectively. The noise generated from this procedure is fully random and uncorrelated.

The illustration of the accuracy of the trained model (in
percentage error) tested using the true values of the
physical parameters ($\theta$) with different noise levels is
shown in Table 3. We follow the procedure proposed by
\cite{kadeethum2020pinn} as we average our results shown
in Table \ref{tab:3} over six realizations (different initializations of
W and b). As expected, this table illustrates that when the
noise level goes up, the model accuracy decreases. The
batch size of 8 or 32 provides the best accuracy.

\begin{table}[htbp]
  \centering
  \caption{Illustration of the accuracy of the trained model (in
percentage error) tested using the true values of the physical
parameters ($\theta_{1}, \theta_{2}, \theta_{3}, \theta_{4},$ and $\theta_{5}$) with different noise level}
    \begin{tabular}{|c|c|c|c|c|c|}
    \hline
    \multicolumn{2}{|c|}{\multirow{2}[4]{*}{batch size}} & \multicolumn{4}{c|}{Noise} \\
\cline{3-6}    \multicolumn{2}{|c|}{} & 0\%   & 1\%   & 5\%   & 10\% \\
    \hline
    \hline
    \multirow{4}[8]{*}{$\theta_1$} & 8     & \cellcolor[rgb]{ .388,  .745,  .482}0.13 & \cellcolor[rgb]{ .388,  .745,  .482}0.17 & \cellcolor[rgb]{ .424,  .757,  .514}0.56 & \cellcolor[rgb]{ .463,  .773,  .545}0.97 \\
\cline{2-6}          & 32    & \cellcolor[rgb]{ .404,  .749,  .494}0.33 & \cellcolor[rgb]{ .459,  .773,  .541}0.94 & \cellcolor[rgb]{ .639,  .847,  .698}2.99 & \cellcolor[rgb]{ .776,  .902,  .82}4.54 \\
\cline{2-6}          & 128   & \cellcolor[rgb]{ .988,  .933,  .945}12.43 & \cellcolor[rgb]{ .988,  .929,  .941}12.83 & \cellcolor[rgb]{ .988,  .945,  .957}11.36 & \cellcolor[rgb]{ .988,  .965,  .976}9.23 \\
\cline{2-6}          & full-batch & \cellcolor[rgb]{ .976,  .416,  .424}64.10 & \cellcolor[rgb]{ .976,  .416,  .424}64.11 & \cellcolor[rgb]{ .973,  .412,  .42}64.20 & \cellcolor[rgb]{ .976,  .416,  .424}64.12 \\
    \hline
    \multirow{4}[8]{*}{$\theta_2$} & 8     & \cellcolor[rgb]{ .573,  .82,  .643}13.11 & \cellcolor[rgb]{ .62,  .839,  .682}14.70 & \cellcolor[rgb]{ .988,  .929,  .941}29.13 & \cellcolor[rgb]{ .984,  .78,  .788}34.62 \\
\cline{2-6}          & 32    & \cellcolor[rgb]{ .388,  .745,  .482}6.82 & \cellcolor[rgb]{ .396,  .749,  .49}7.20 & \cellcolor[rgb]{ .553,  .812,  .624}12.40 & \cellcolor[rgb]{ .922,  .961,  .945}24.87 \\
\cline{2-6}          & 128   & \cellcolor[rgb]{ .882,  .945,  .91}23.51 & \cellcolor[rgb]{ .922,  .961,  .941}24.85 & \cellcolor[rgb]{ .988,  .89,  .902}30.62 & \cellcolor[rgb]{ .984,  .725,  .733}36.64 \\
\cline{2-6}          & full-batch & \cellcolor[rgb]{ .976,  .42,  .427}47.60 & \cellcolor[rgb]{ .976,  .42,  .427}47.67 & \cellcolor[rgb]{ .976,  .42,  .427}47.70 & \cellcolor[rgb]{ .973,  .412,  .42}47.86 \\
    \hline
    \multirow{4}[8]{*}{$\theta_3$} & 8     & \cellcolor[rgb]{ .478,  .78,  .561}2.06 & \cellcolor[rgb]{ .514,  .796,  .592}2.41 & \cellcolor[rgb]{ .773,  .898,  .812}4.84 & \cellcolor[rgb]{ .894,  .949,  .918}6.01 \\
\cline{2-6}          & 32    & \cellcolor[rgb]{ .388,  .745,  .482}1.18 & \cellcolor[rgb]{ .439,  .765,  .525}1.67 & \cellcolor[rgb]{ .643,  .847,  .702}3.63 & \cellcolor[rgb]{ .788,  .906,  .827}4.99 \\
\cline{2-6}          & 128   & \cellcolor[rgb]{ .988,  .98,  .992}7.74 & \cellcolor[rgb]{ .988,  .98,  .992}8.12 & \cellcolor[rgb]{ .988,  .98,  .992}7.94 & \cellcolor[rgb]{ .988,  .976,  .988}8.26 \\
\cline{2-6}          & full-batch & \cellcolor[rgb]{ .973,  .412,  .42}69.76 & \cellcolor[rgb]{ .976,  .416,  .424}69.35 & \cellcolor[rgb]{ .976,  .42,  .427}69.10 & \cellcolor[rgb]{ .976,  .416,  .424}69.54 \\
    \hline
    \multirow{4}[8]{*}{$\theta_4$} & 8     & \cellcolor[rgb]{ .396,  .745,  .486}0.53 & \cellcolor[rgb]{ .4,  .749,  .49}0.58 & \cellcolor[rgb]{ .388,  .745,  .482}0.43 & \cellcolor[rgb]{ .412,  .753,  .502}0.72 \\
\cline{2-6}          & 32    & \cellcolor[rgb]{ .412,  .753,  .502}0.72 & \cellcolor[rgb]{ .431,  .761,  .518}0.97 & \cellcolor[rgb]{ .529,  .8,  .604}2.19 & \cellcolor[rgb]{ .643,  .847,  .702}3.59 \\
\cline{2-6}          & 128   & \cellcolor[rgb]{ .988,  .867,  .878}13.25 & \cellcolor[rgb]{ .988,  .859,  .871}13.60 & \cellcolor[rgb]{ .988,  .855,  .867}13.67 & \cellcolor[rgb]{ .988,  .894,  .906}11.99 \\
\cline{2-6}          & full-batch & \cellcolor[rgb]{ .976,  .416,  .424}32.99 & \cellcolor[rgb]{ .976,  .416,  .424}32.97 & \cellcolor[rgb]{ .976,  .416,  .424}33.09 & \cellcolor[rgb]{ .973,  .412,  .42}33.13 \\
    \hline
    \multirow{4}[8]{*}{$\theta_5$} & 8     & \cellcolor[rgb]{ .455,  .773,  .541}7.74 & \cellcolor[rgb]{ .482,  .784,  .565}8.66 & \cellcolor[rgb]{ .714,  .875,  .761}16.24 & \cellcolor[rgb]{ .812,  .918,  .847}19.58 \\
\cline{2-6}          & 32    & \cellcolor[rgb]{ .388,  .745,  .482}5.46 & \cellcolor[rgb]{ .404,  .749,  .494}6.01 & \cellcolor[rgb]{ .541,  .804,  .612}10.56 & \cellcolor[rgb]{ .804,  .914,  .839}19.23 \\
\cline{2-6}          & 128   & \cellcolor[rgb]{ .988,  .941,  .949}31.69 & \cellcolor[rgb]{ .988,  .929,  .941}32.81 & \cellcolor[rgb]{ .988,  .929,  .937}33.14 & \cellcolor[rgb]{ .988,  .945,  .957}31.03 \\
\cline{2-6}          & full-batch & \cellcolor[rgb]{ .976,  .416,  .424}97.49 & \cellcolor[rgb]{ .976,  .416,  .424}97.45 & \cellcolor[rgb]{ .973,  .412,  .42}97.50 & \cellcolor[rgb]{ .976,  .416,  .424}97.46 \\
    \hline
    \end{tabular}%
  \label{tab:3}%
\end{table}%

\section{General discussion}
From the above findings, we see that batch training may enhance the accuracy of the PINN model for physical parameter estimation. The possible applications of this model range from estimating rock properties, e.g., porosity, permeability, bulk modulus, Biot’s coefficient, and Poisson ratio, from a lab setting to the field scale when the measurement fields of pressure and displacement are available. There are still main challenges to be addressed in the field application since the pressure or displacement information can only be obtained at some spatial coordinates where wells are drilled. The first one involves an incomplete dataset, i.e., displacements or pressure data points are not available at the same spatial and temporal coordinates. Moreover, a biased sampling situation is needed to investigate to represent the case where we only measure data points at specific wells.

Even though we only apply PINN with batch training for the Biot’s system in this study, the PINN approach has been successfully applied to other (multi)physics problems such as fluid dynamics \cite{raissi2017inferring, raissi2019physics}, solid mechanics \cite{haghighat2020deep}, and (multi)phase flow in porous media \cite{wang2020efficient}. PINN, in general, requires much less training data comparing to the traditional artificial neural network \cite{raissi2019physics}. It may be worth applying to other types of physics or enhancing conventional numerical methods, as mentioned in \cite{baker2019workshop}.

\section{Conclusion}

This paper extends the physics-informed neural network for solving an inverse problem of nonlinear Biot’s equations presented in \cite{kadeethum2020pinn} to batch training. Our results show that batch training may provide better accuracy for the predicted values of the neural network and the estimated physical parameters (Section 4). One, however, should be aware of the tradeoff between accuracy and training time when selecting the optimal batch size. The smaller the batch size, the higher the training time (Section 5). On the other hand, if the batch size is too large, the model accuracy is decreased (Section 5). The learning rate also plays an important role and accordingly is a vital hyperparameter (Section 6). We also note that since the PINN model is stochastic, the final results varied with different initializations. The general trend, however, remains the same (Section 7). The batch training (batch size = 8 or 32) also tolerates noisy data and still predicts the physical parameters with a percentage error of less than 25 \% with a noise level of 10 \% (Section 8).  In a future study, transfer learning and batch normalization will be explored as the means of reducing the batch training time.

\section*{Acknowledgments}
The research leading to these results has received funding from the Danish Hydrocarbon Research and Technology Centre under the Advanced Water Flooding program.

\bibliographystyle{unsrt}  
\bibliography{references.bib}

\end{document}